\documentclass[a4paper, twocolumn]{article}

\usepackage[margin=1in]{geometry} 

\usepackage[utf8]{inputenc}
\usepackage{algorithm, algpseudocode} 
\usepackage{mathrsfs} 

\usepackage{indentfirst}
\usepackage{bm}
\usepackage{cite}
\usepackage{amsmath,amssymb,amsfonts}
\usepackage{graphicx}
\usepackage{textcomp}
\usepackage{xcolor}
\usepackage[acronym]{glossaries}

\usepackage{booktabs}
\usepackage{enumerate}
\usepackage{multicol}
\usepackage{multirow}
\usepackage{array}
\usepackage[flushleft]{threeparttable}
\usepackage{transparent}
\usepackage{enumitem}

\usepackage{longtable}
\usepackage{afterpage}
\usepackage{pdflscape}
\usepackage{capt-of}
\usepackage{threeparttablex}
\usepackage{chngpage}
\usepackage{colortbl}
\usepackage{caption}
\usepackage{multirow}
\usepackage{adjustbox}
\usepackage{textgreek}

\usepackage{xr-hyper}
\usepackage[colorlinks=true,linkcolor=black,anchorcolor=black,citecolor=black,filecolor=black,menucolor=black,runcolor=black,urlcolor=black]{hyperref}

\newcolumntype{C}[1]{>{\centering\arraybackslash}p{#1}}
\makeatletter
\def\bstctlcite{\@ifnextchar[{\@bstctlcite}{\@bstctlcite[@auxout]}}
\def\@bstctlcite[#1]#2{\@bsphack
  \@for\@citeb:=#2\do{%
    \edef\@citeb{\expandafter\@firstofone\@citeb}%
    \if@filesw\immediate\write\csname #1\endcsname{\string\citation{\@citeb}}\fi}%
  \@esphack}
\makeatother 

\newacronym{SeFEF}{SeFEF}{Seizure Forecasting Evaluation Framework}

\newacronym{IT}{IT}{Instituto de Telecomunicações}
\newacronym{FCT}{FCT}{Fundação para a Ciência e Tecnologia}
\newacronym{MSG}{MSG}{My Seizure Gauge}
\newacronym{MSG2022}{MSG2022}{My Seizure Gauge Forecasting Challenge 2022}
\newacronym{ULSSM}{ULSSM}{Unidade Local de Saúde Santa Maria}

\newacronym{CV}{CV}{cross-validation}
\newacronym{GMM}{GMM}{Gaussian mixture model}
\newacronym{LR}{LR}{logistic regression}
\newacronym{ML}{ML}{machine learning}
\newacronym{SI}{SI}{synchronization index}
\newacronym{TSCV}{TSCV}{time series cross-validation}

\newacronym{CDF}{CDF}{cumulative distribution function}
\newacronym{PDF}{PDF}{probability density function}

\newacronym{AUC}{AUC}{area under the curve}
\newacronym{AUCSenTiw}{AUC(Sen,Tiw)}{area under the curve of sensitivity vs time in warning}
\newacronym{BS}{BS}{Brier score}
\newacronym{BSS}{BSS}{Brier skill score}
\newacronym{CC}{CC}{calibration curve}
\newacronym{GSS}{GSS}{geometric mean of sensitivity and specificity}
\newacronym{IoC}{IoC}{improvement over chance}
\newacronym{FP}{FP}{false positive}
\newacronym{FPR}{FPR}{false positive rate}
\newacronym{FN}{FN}{false negative}
\newacronym{MAPE}{MAPE}{mean absolute error}
\newacronym{MRPE}{MRPE}{relative percentage error}
\newacronym{OR}{OR}{odds ratio}
\newacronym{RR}{RR}{risk ratio}
\newacronym{Sen}{Sen}{sensitivity}
\newacronym{Spe}{Spe}{specificity}
\newacronym{TiW}{TiW}{time in warning}
\newacronym{TN}{TN}{true negative}
\newacronym{TP}{TP}{true positive}
\newacronym{WBC}{WBC}{within-bin covariance}
\newacronym{WBV}{WBV}{within-bin variance}

\newacronym{ACC}{ACC}{accelerometer}
\newacronym{ANS}{ANS}{autonomic nervous system}
\newacronym{bpm}{bpm}{beats-per-minute}
\newacronym{BVP}{BVP}{blood volume pulse}
\newacronym{CDE}{CDE}{cyclic distribution of events}
\newacronym{ECG}{ECG}{electrocardiography}
\newacronym{EDA}{EDA}{electrodermal activity}
\newacronym{EEG}{EEG}{electroencephalography}
\newacronym{EMG}{EMG}{electromyography}
\newacronym{EOG}{EOG}{electrooculography}
\newacronym{GYR}{GYR}{gyroscope}
\newacronym{HR}{HR}{heart rate}
\newacronym{HRV}{HRV}{heart rate variability}
\newacronym{IEA}{IEA}{inter-ictal epileptiform activity}
\newacronym{iEEG}{iEEG}{intracranial electroencephalography}
\newacronym{LUX}{LUX}{light}
\newacronym{MAG}{MAG}{magnetic field-based}
\newacronym{PZT}{PZT}{piezoelectric respiration}
\newacronym{PCG}{PCG}{phonocardiography}
\newacronym{PPG}{PPG}{photopletysmography}
\newacronym{RESP}{RESP}{respiration}
\newacronym{sEEG}{sEEG}{surface electroencephalography}
\newacronym{SMPS}{SMPS}{surrogate measures of the pre-ictal state}
\newacronym{SpO2}{SpO2}{blood oxygen saturation}
\newacronym{TEMP}{TEMP}{temperature}
\newacronym{ToD}{ToD}{time-of-day}
\newacronym{vEEG}{vEEG}{video-electroencephalography}

\newacronym{FAS}{FAS}{focal aware seizure}
\newacronym{FAMS}{FAMS}{focal aware motor seizure}
\newacronym{FANMS}{FANMS}{focal aware nonmotor seizure}
\newacronym{FBTCS}{FBTCS}{focal to bilateral tonic-clonic seizure}
\newacronym{FIAS}{FIAS}{focal impaired awareness seizure}
\newacronym{FIAMS}{FIAMS}{focal impaired awareness motor seizure}
\newacronym{FIANMS}{FIANMS}{focal impaired awareness nonmotor seizure}
\newacronym{FUAMS}{FUAMS}{focal unknown awareness motor seizure}
\newacronym{FUANMS}{FUANMS}{focal unknown awareness nonmotor seizure}
\newacronym{FUAS}{FUAS}{focal unknown awareness seizure}
\newacronym{GTCS}{GTCS}{generalized tonic-clonic seizure}
\newacronym{GMS}{GMS}{generalized motor seizure}
\newacronym{GNMS}{GNMS}{generalized nonmotor (absence) seizure}
\newacronym{UOMS}{UOMS}{unknown onset motor seizure}
\newacronym{UONMS}{UONMS}{unknown onset nonmotor seizure}

\title{SeFEF: A Seizure Forecasting Evaluation Framework}

\author{\parbox{\linewidth}{\centering Ana Sofia Carmo$^{1,2*}$, Lourenço Abrunhosa Rodrigues$^3$, Ana Rita Peralta$^{4,5}$, Ana Fred$^{1,2}$, Carla Bentes$^{4,5}$ and Hugo Plácido da Silva$^{1,2,6}$}}

\date{
    $^1$Dept. Bioengineering, Instituto Superior Técnico, Universidade de Lisboa, Lisboa, Portugal\\ 
    $^2$Instituto de Telecomunicações, Lisboa, Portugal\\
    $^3$Cognitive Systems Lab, FB03 - Mathematics and Informatics, Universität Bremen, Bremen, Germany\\
    $^4$Neurophysiology Monitoring Unit - EEG/Sleep Laboratory, Hospital de Santa Maria, Unidade Local de Saúde Santa Maria, Lisboa, Portugal\\%
    $^5$Centro de Estudos Egas Moniz. Faculdade de Medicina da Universidade de Lisboa, Lisboa, Portugal\\
    $^6$LUMLIS - The Lisbon ELLIS Unit $|$ European Laboratory for Learning and Intelligent Systems, Portugal\\[1ex]
    \texttt{*ana.sofia.carmo@tecnico.ulisboa.pt}\\[2ex]%
}

\begin{document}
\onecolumn
\maketitle

\begin{abstract}
  \textbf{Background and Objective:} The unpredictability of seizures remains a major challenge for people with epilepsy, affecting their daily lives and overall well-being. Seizure forecasting offers a potential solution, but the absence of standardized protocols for algorithm development and evaluation makes it difficult to compare different approaches directly. This lack of consistency slows progress in the field and limits the clinical translation of forecasting models. In this work, we introduce a Python-based framework aimed at streamlining the development, assessment, and documentation of individualized seizure forecasting algorithms.

  \textbf{Methods:} The framework automates data labeling, cross-validation splitting, forecast post-processing, performance evaluation, and reporting. It supports various forecast horizons and includes a model card that documents implementation details, training and evaluation settings, and performance metrics. Three different models were implemented as a proof-of-concept, to demonstrate the framework's versatility across diverse input data and algorithmic approaches. The models leveraged features extracted from time series data, seizure periodicity, and a combination of these. Model performance was assessed using time series cross-validation, and key metrics such as area under the curve of sensitivity vs time in warning (AUC(Sen,Tiw)), Brier score (BS), and Brier skill score (BSS).

  \textbf{Results:} The proposed framework streamlines seizure forecasting model development by providing a standardized, reproducible evaluation pipeline. Implementation of the three models was successful, demonstrating the flexibility of the framework. While the results illustrate its versatility across diverse forecasting approaches and input data, they also emphasize the importance of careful model interpretation due to variations in probability scaling, calibration, and subject-specific differences. Although formal usability metrics were not recorded, empirical observations suggest that using the framework reduced development time and helped maintain methodological consistency, minimizing unintentional variations that could affect the comparability of different approaches. 

  \textbf{Conclusions:} As a proof-of-concept, this validation is inherently limited, relying on a single-user experiment without statistical analyses or replication across independent datasets. Regardless, at this stage, our objective is to make the framework publicly available to foster community engagement, facilitate experimentation, and gather feedback. In the long term, we aim to contribute to the establishment of a consensus on a standardized methodology for the development and validation of seizure forecasting algorithms in people with epilepsy.

  \textbf{Keywords:} Seizure Forecasting; Python Framework; Machine Learning; Benchmarking; Time Series Analysis; Biomedical Signal Processing 
\end{abstract}

\section{Introduction}

As a result of a 2016 community survey \cite{EpilepsyFoundation20162016Community}, the Epilepsy Foundation recognized seizure forecasting as a potential solution to the nefarious effects of the unpredictability of seizures. Since then, several automated seizure forecasting algorithms have been proposed in literature, with promising results.

Several reports of non-physiological factors that appear to be linked to seizure occurrence \cite{Grzeskowiak2021SeizureForecasting,Dell2021SeizureLikelihood,Doherty2007AtmosphericPressure}, as well as of physiological manifestations preceding the onset of seizures \cite{Hubbard2021ChallengingPath,Vieluf2021TwentyfourhourPatterns,Leal2023UnsupervisedEEG}, have encouraged the use of a diverse range of data as input to these automated algorithms \cite{Carmo2024AutomatedAlgorithms}. Proposed methodologies to analyze data and report performances have also been shown to be notoriously varied in terms of horizon chosen for the forecast, algorithmic approaches used, and how authors choose to report performance \cite{Carmo2024AutomatedAlgorithms}.

Such lack of standardization in protocols for developing and testing algorithms inhibits direct comparison between state-of-the-art approaches, both hindering progress in the field and leaving the epilepsy community without an accurate assessment of current seizure forecasting capabilities \cite{Carmo2024AutomatedAlgorithms,Dan2024SzCORESeizurea}. As recently proposed in a systematic literature review \cite{Carmo2024AutomatedAlgorithms}, standardized protocols for the development, evaluation, and reporting of new proposed approaches are key to tackle this issue.

In this paper we propose to facilitate adherence to the guidelines recommended in \cite{Carmo2024AutomatedAlgorithms} through the \gls{SeFEF}\footnote{The framework is open-source and available on GitHub \href{https://github.com/anascacais/SeFEF}{github.com/anascacais/SeFEF}, and can be installed via \href{https://pypi.org/project/sefef/}{PyPI}.}, a Python-based framework that standardizes the development, evaluation, and reporting of individualized algorithms for seizure forecasting. The proposed framework also aims to decrease development time and minimize implementation errors, by automating key procedures within data preparation, training/testing, and computation of evaluation metrics. These features strive to provide consistent and reliable assessments of forecasting algorithms, streamline the development process, and aid in addressing research questions effectively.

As a proof-of-concept, the framework was applied to 3 diverse algorithmic approaches using a benchmark dataset.

\section{Methods}

The task of seizure forecasting can be defined as the prediction of the probability of a seizure occurring within a specific time period after the forecast, called the forecast horizon \cite{Carmo2024AutomatedAlgorithms}. When addressing this task, several steps are implied within model development and evaluation.

\subsection{Standardized Forecasting Framework}

\begin{figure}[!ht]
  \centering
  \includegraphics[width=1\linewidth]{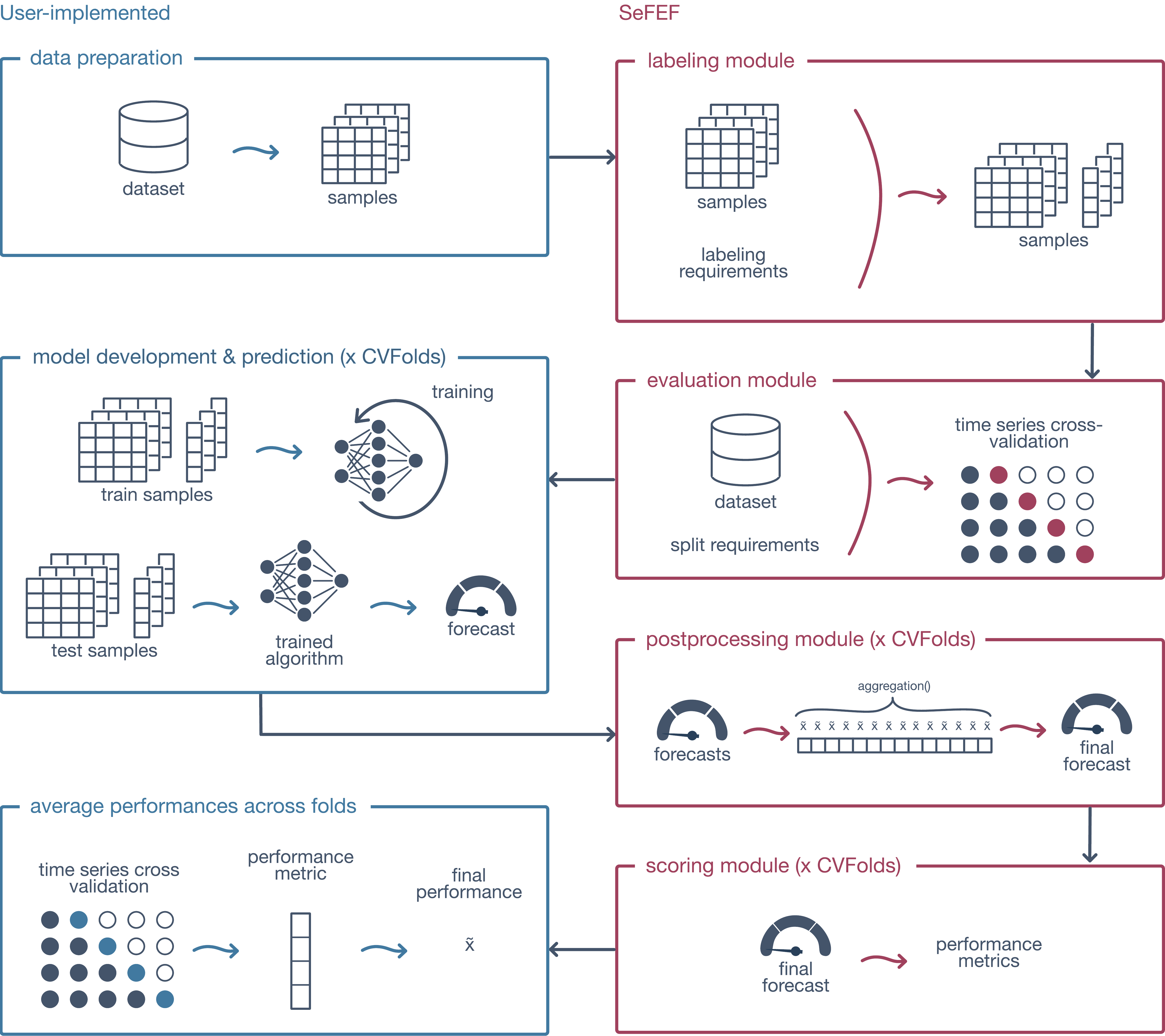}
  \caption{Illustration of the steps implied in the task of seizure forecasting. The rose boxes (on the right) correspond to steps that are automated by \acrshort{SeFEF}, while the blue boxes (on the left) are implemented by the user. \footnotesize Acronyms: \textbf{CVFolds}: \# of folds in \acrlong{CV}.}
  \label{fig:SeFEF}
\end{figure}

Our proposed conceptual for a standardized forecasting framework is illustrated in Figure \ref{fig:SeFEF} and is characterized by the following steps: data preparation, labeling, \acrlong{TSCV}, model development, prediction, forecast post-processing, and performance scoring.

\subsubsection*{Data Preparation}

Preparation of data is handled by the user, supporting full customization of data processing procedures and multiscale modeling (i.e., handling data at different time resolutions). The framework expects a HDF5 file with two datasets: one with the samples and another with the timestamps corresponding to the start time of the samples. Two distinct examples of data preparation in the context of seizure forecasting are provided in Section \ref{sec:data_preparation}.

\subsubsection*{Labeling}
Given the probabilistic framework of seizure forecasting, contrarily to seizure prediction, the output is an equally-spaced and continuous measure of probability \cite{Carmo2024AutomatedAlgorithms}. However, binary labeling is still typically used during model training (inter-ictal/pre-ictal), in which the pre-ictal duration and onset setback (i.e., the time before the onset of a seizure) are determined according to empirical or clinical knowledge. This concept, analogous to seizure prediction, aims to train the model to recognize the patterns and dynamics that may be associated to the future occurrence of a seizure. 

Therefore, each individual sample is labeled according to the chosen \textit{pre-ictal duration} and \textit{onset setback}.

\subsubsection*{Time Series Cross-Validation}
\label{sec:evaluation}

\begin{figure}[!ht]
  \centering
  \includegraphics[width=1.\linewidth]{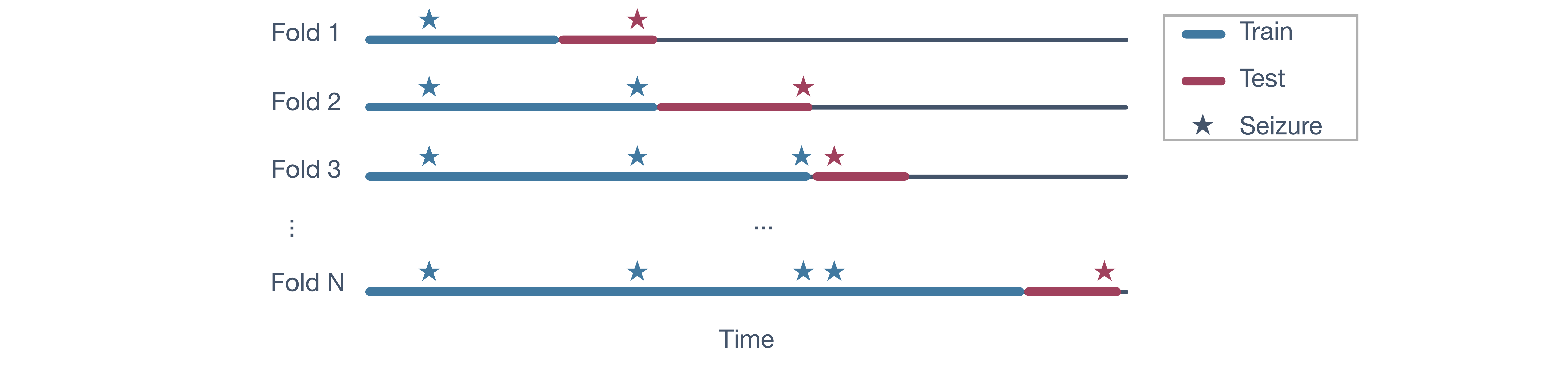}
  \caption{Illustration of \acrlong{TSCV} applied to the context of seizure forecasting.}
  \label{fig:TSCV}
\end{figure}

In intrasubject prediction tasks, a particular form of \gls{CV} is often used, namely the \gls{TSCV} \cite{Hyndman2021Chapter510}. As illustrated in Figure \ref{fig:TSCV}, \gls{TSCV} ensures that the data used to train the model consist only in observations prior to the ones used to evaluate model performance (also known as pseudo-prospective evaluation). This evaluation methodology respects the chronological order of events, mimicking the real-world use case of prospective evaluation \cite{Carmo2024AutomatedAlgorithms}, and allowing for adequate handling of non-stationarity in time series data \cite{Bergmeir2018NoteValidity}. 

\subsubsection*{Model Development and Prediction}

Model development is handled by the user, including training, internal validation, hyperparameter tuning, and threshold optimization, allowing for full flexibility in approach complexity, hierarchical modeling, and custom methodologies, while ensuring compatibility with the framework's standardized evaluation pipeline. Three distinct algorithmic approaches are presented in Section \ref{sec:model_training}. 

\subsubsection*{Forecast Post-Processing}

The post-processing methodology implemented in \gls{SeFEF} is inspired by one described in the literature \cite{Nasseri2021AmbulatorySeizure,Attia2021SeizureForecasting,Viana2023SeizureForecasting}. For each time period, with a duration equal to the forecast horizon, mean predicted probabilities are computed for groups of consecutive, non-overlapping samples. The maximum probability within each period is then selected. Consequently, a seizure forecast is issued at the start of each new period, based on predictions made during the preceding period. This methodology is formalized through Equation (\ref{eq:postprocessing}).

\begin{equation}
  \label{eq:postprocessing}
  \hat{p}_t = \max \left( \frac{1}{N} \sum_{i=1}^{N} p_{t-i} \right), \quad t \in T_H
\end{equation}

\noindent
where $\hat{p}_t$ is the final seizure forecast at time $t$, $T_H$ represents the discrete set of times points that indicate the start of each forecast horizon, $p_{t-i}$ are the predicted probabilities for individual samples within the period preceding $t$, and $N$ is the number of samples within an averaging window.

\begin{figure}[!ht]
    \begin{center}
        \includegraphics[width=.8\linewidth]{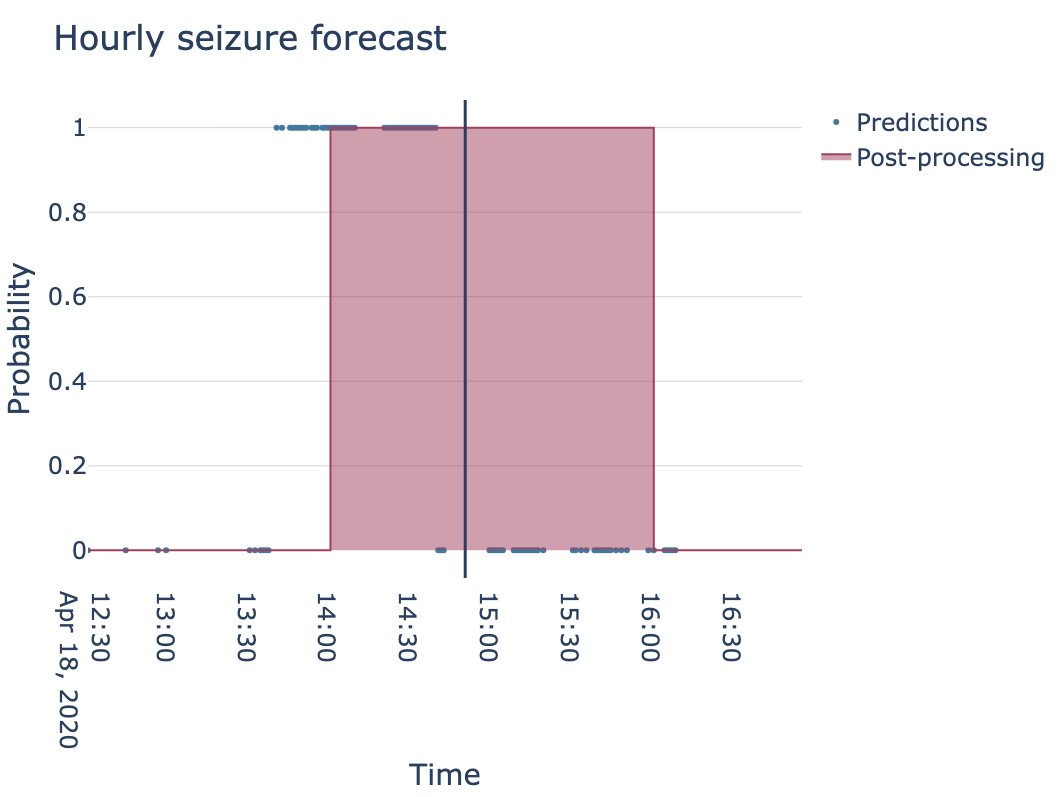}%
    \end{center}
    \caption{Effect of post-processing methodology using the optimal prediction case (i.e., using the actual labels as predicted probabilities). A vertical black bar identifies the onset of the seizure.}
    \label{fig:optimal_prediction_postprocess}
\end{figure}

Figure \ref{fig:optimal_prediction_postprocess} illustrates this methodology, where mean predicted probabilities were computed every 5 minutes (i.e., over five consecutive 60-second samples), and a forecast was issued at the start of each clock-hour. The image demonstrates the effect of this post-processing in the \textit{optimal} prediction case, where actual labels are used as predicted probabilities. 

\subsubsection*{Performance Scoring}

Probabilistic forecasts have been historically evaluated through deterministic metrics \cite{Carmo2024AutomatedAlgorithms}, which require the conversion of the predicted probability into high/low likelihood states, and often rely on threshold optimization (high/low likelihood thresholds). The categorical likelihood states are leveraged to define the classical concepts of \glspl{TP}, \glspl{FP}, and \glspl{FN}, which can then be used to compute standard performance metrics such as \gls{Sen} and \gls{FPR}. 

Despite the commonality of deterministic metrics, some authors have started evaluating their proposed approaches through probabilistic measures of performance \cite{Carmo2024AutomatedAlgorithms}, borrowed from the domain knowledge of weather forecasting. As proposed by Mason \cite{Mason2015GuidanceVerification}, the quality of a forecast is complex and multidimensional. It includes attributes such as how strongly the outcome is conditioned on the forecast (i.e., resolution), if the confidence of the forecast is in agreement with the observed rate of the corresponding outcome (i.e., reliability), and how much better the forecast is when compared to a reference (i.e., skill). 

Given this duality of deterministic and probabilistic performance evaluation, and also reflecting the trends in the literature, \gls{SeFEF} provides a set of both deterministic and probabilistic scoring metrics. A comprehensive description of these metrics is provided in Appendix \ref{app:performance}.

\subsection{Implementation of the Framework}

\gls{SeFEF} was implemented in Python and follows a modular design, where each module can be used independently. Each module automates one of the steps defined above, consisting on \verb|labeling|, \verb|evaluation| (which addresses \gls{TSCV}), \verb|postprocessing|, and \verb|scoring|. Additionally, two auxiliary modules, \verb|modelcard| and \verb|visualization|, are included, providing support for reporting and visualization.

The framework is open-source and available on GitHub \href{https://github.com/anascacais/SeFEF}{github.com/anascacais/SeFEF}, and can be installed via PyPI: \verb|pip install sefef|.

\subsection{Preparation of the Proof-of-Concept}
\label{sec:proof_of_concept}

The \gls{MSG2022} dataset was used as a proof-of-concept of \gls{SeFEF}, providing 2 distinct types of input data to the framework. 

\subsubsection*{MSG2022 Dataset}

The \gls*{MSG2022} dataset\footnote{The \acrfull*{MSG2022} dataset was shared at \url{https://eval.ai/web/challenges/challenge-page/1693/overview} (Online, last accessed July 2024), but the data repository was archived on April 2024.} was launched under the \gls*{MSG} project and comprises continuous, long-term, wearable data from 6 patients, spanning approximately 993 days in total. The data made publicly available is a subset of the data described in \cite{Nasseri2021AmbulatorySeizure}. The \gls*{MSG2022} data consists of annotations of seizure onsets, as well as 8 channels with continuous physiological recordings (derived from the recordings of an Empatica E4 device, Empatica Inc., Boston MA), namely 3-axial \gls*{ACC}, \gls*{ACC} magnitude, \gls*{BVP}, \gls*{EDA}, \gls*{HR}, and \gls*{TEMP}. 

\subsubsection*{Types of Input Data}

\begin{figure}[!ht]
  \centering
  \begin{minipage}{.33\textwidth}
      \centering
      \includegraphics[width=1\linewidth]{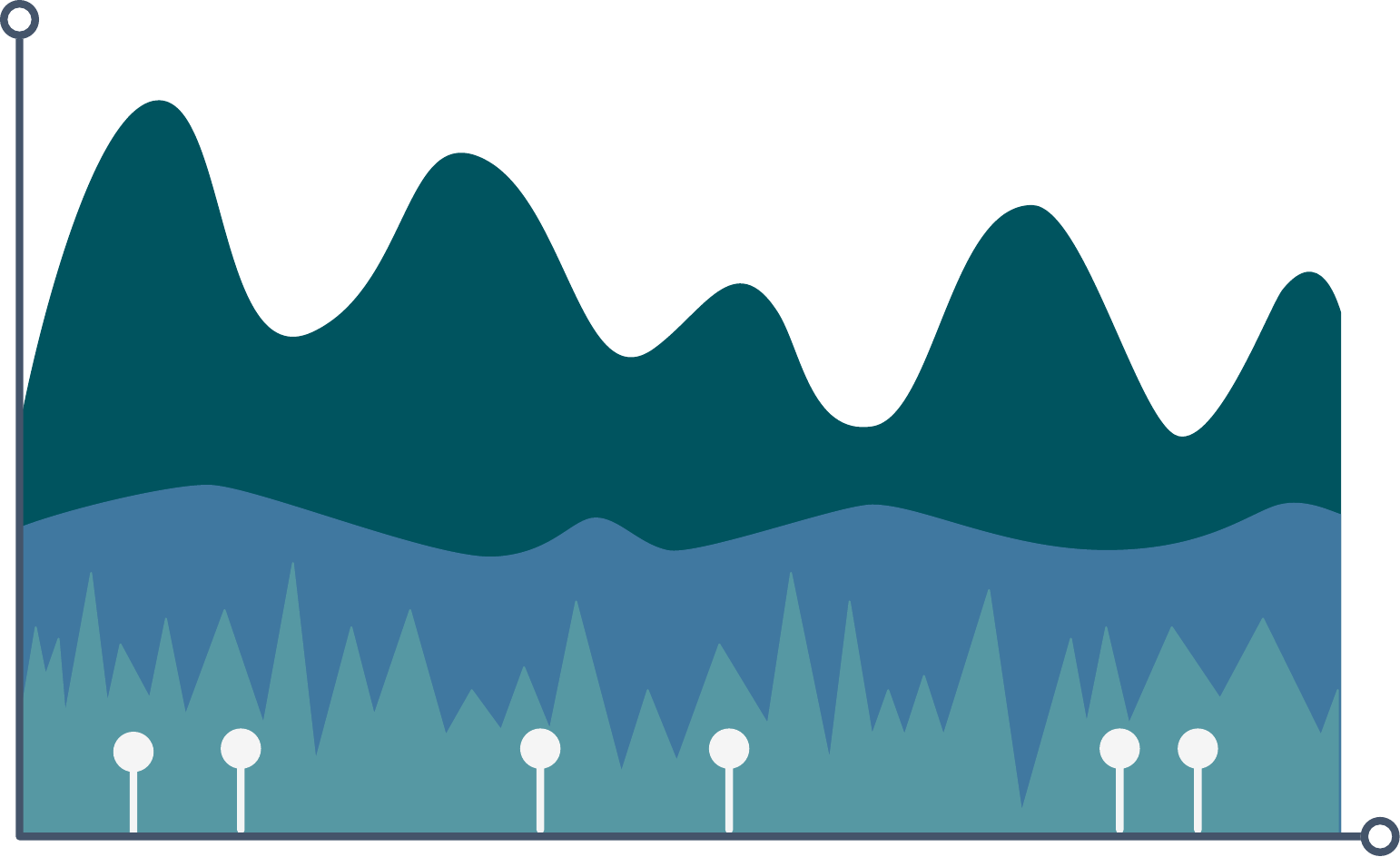}
      (a)
  \end{minipage}%
  \begin{minipage}{.33\textwidth}
    \centering
    \includegraphics[width=1\linewidth]{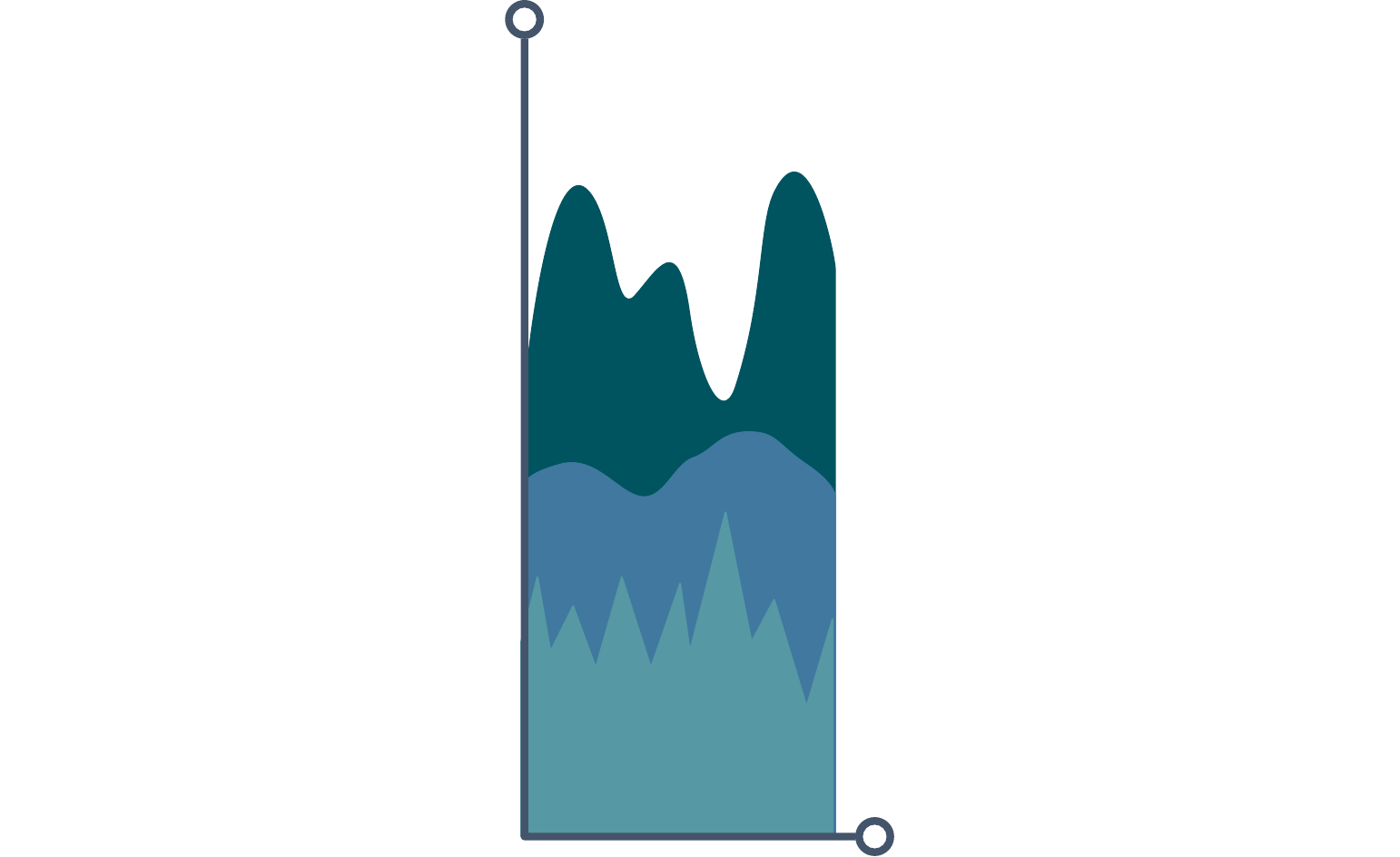}
    (b)
\end{minipage}%
\begin{minipage}{.33\textwidth}
  \centering
  \includegraphics[width=1\linewidth]{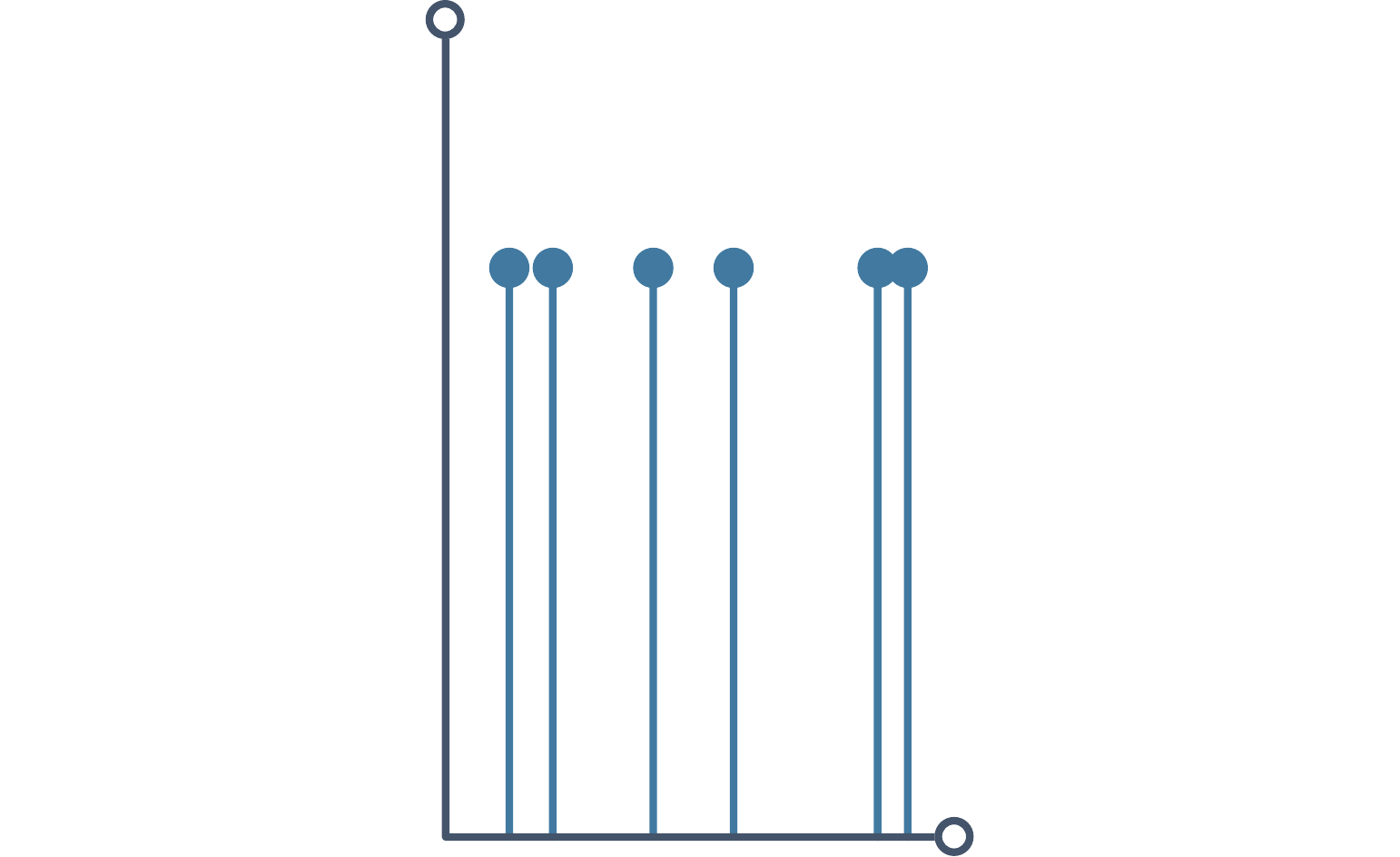}
  (c)
\end{minipage}%
\caption{Illustration of 2 types of input data that can be represented from a multimodal and long-term dataset (as illustrated in (a)): (b) continuous, multimodal time series; and (c) sporadic, timestamp-only, seizure onset records.}
\label{fig:datasets}
\end{figure}

The multimodal and long-term nature of this dataset enables the representation of two distinct input data types (illustrated in Figure \ref{fig:datasets}): continuous multimodal time series; and sporadic, timestamp-only, seizure onset records. This structure provides a valuable foundation for replicating diverse algorithmic approaches proposed in the literature. Continuous time series data can be leveraged for feature extraction, feeding traditional machine learning models, or directly processed by deep learning architectures. In contrast, seizure onset timestamps facilitate the modeling of intrinsic periodicities in seizure occurrence.

\section{Proof-of-Concept}

This section presents the results of three different algorithmic approaches implemented using \gls{SeFEF}, demonstrating its versatility in handling diverse input data types and forecasting methodologies. The framework was used throughout the whole model development process, which included automation of cross-validation splits, labeling of data, post-processing of forecasts, computation of deterministic and probabilistic evaluation metrics, and visualization of results. The primary objective of this evaluation was not to achieve optimal model performance, but to validate that the framework effectively supports diverse data input types as well as multiple seizure forecasting approaches.

Each method was inspired by prior work described in the literature \cite{Karoly2020ForecastingCycles, Karoly2017CircadianProfile}, and was implemented so that they fit within the standardized pipeline provided by \gls{SeFEF}. The approaches differ in their data requirements, preprocessing steps, and forecasting strategies, showcasing the framework's flexibility. Table \ref{tab:summary_approaches} summarizes the algorithmic approaches described below.

\begin{table}[!ht]
  \centering
  \caption{Summary of the algorithmic approaches described in this work. \footnotesize Acronyms: \textbf{\acrshort{LR}:} \acrlong{LR}; \textbf{\acrshort{SI}:} \acrlong{SI}.}
  \label{tab:summary_approaches}
  \begin{tabular}{lp{0.3\linewidth}p{0.4\linewidth}}
    \toprule
    \textbf{Name} & \textbf{Input data} & \textbf{Algorithmic approach} \\
    \toprule
      Von Mises estimator & Timestamps of seizure onsets & Seizure periodicity found through \acrshort{SI}, von Mises distribution fit to data, and probability estimated from distribution. \\\\
      \acrshort{LR} & Time series features & \acrshort{LR} fit to time series features with L2 regularization.\\\\
      Periodicity-aware \acrshort{LR} & Time series features \& timestamps of seizure onsets & \acrshort{LR} with von Mises estimator used as seizure prior.\\
    \bottomrule 
  \end{tabular}
\end{table}

\subsection{Data Preparation}
\label{sec:data_preparation}

Time series data from 4 of the available channels (i.e., \gls*{ACC} magnitude, \gls*{BVP}, \gls*{EDA}, and \gls*{HR}) were segmented into 60-second windows and preprocessed, followed by the extraction of a set of 36 features. A comprehensive description of the preprocessing methodology and feature extraction details is made available in Appendix \ref{app:data}. The final time series dataset consisted on the complete set of 60-second windows (i.e., the samples) extracted from the original data files, each represented by a feature vector and the Unix-timestamp corresponding to the start time of the window. 

The final seizure timestamps  dataset was synthetically generated from seizure onset timestamps. Each sample represented a contiguous time period equal to the forecast horizon (e.g., 1h or 24h). The dataset spanned from the first to the last seizure onset, with an additional sample before the first onset and another after the last onset. Appendix \ref{app:data} provides a visualization of the methodology described.

\subsection{Time Series Cross-Validation}

\begin{figure}[!ht]
  \centering
  \begin{minipage}{.51\textwidth}
      \centering
      \includegraphics[width=1\linewidth]{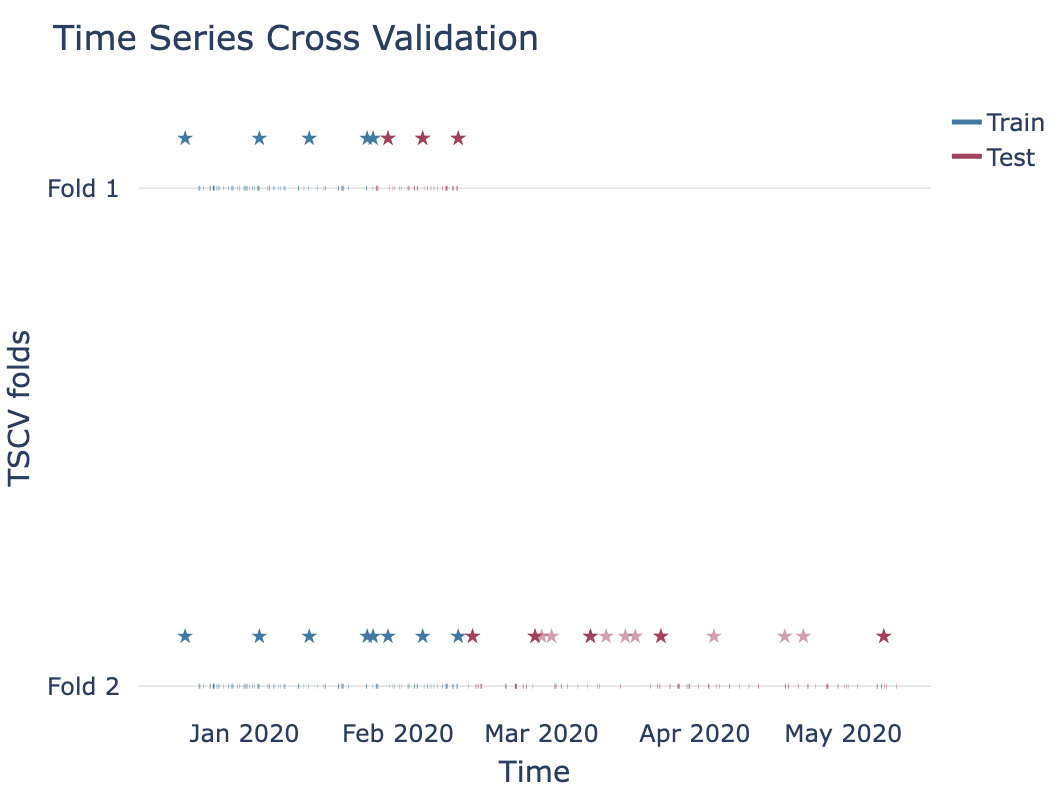}
      (a)
  \end{minipage}%
  \begin{minipage}{.51\textwidth}
    \centering
    \includegraphics[width=1\linewidth]{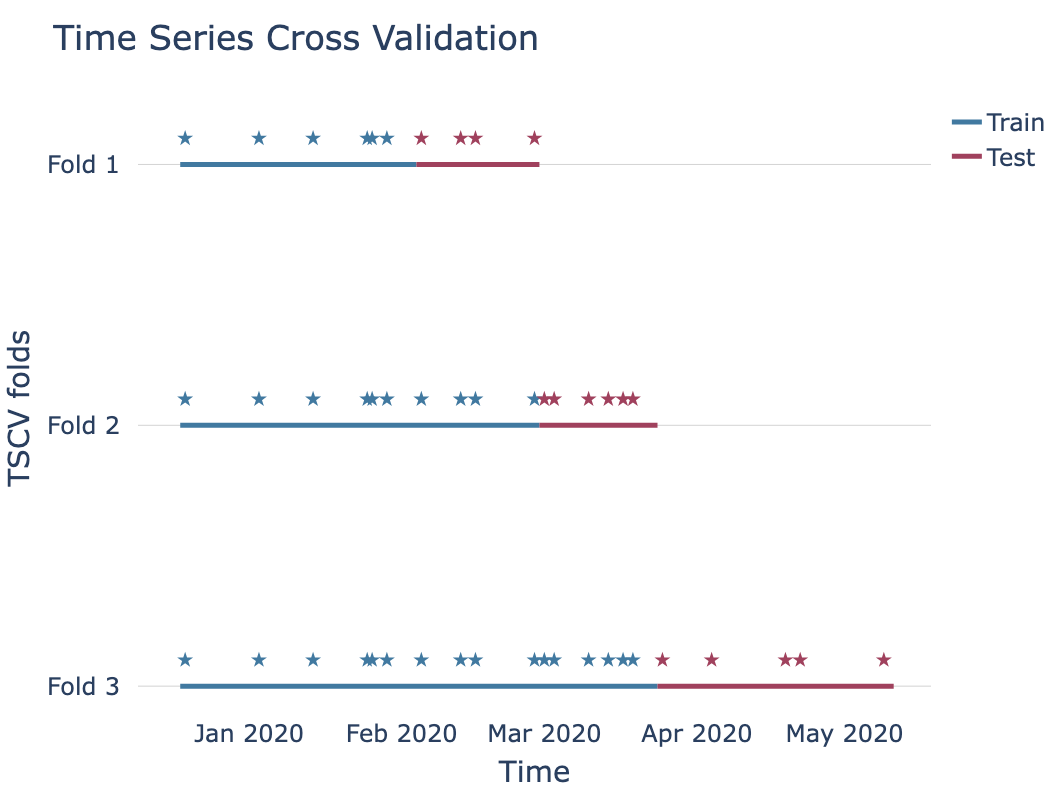}
    (b)
\end{minipage}%
\caption{\Acrlong{TSCV} for (a) time series dataset and (b) seizure timestamps dataset, for patient 1876, achieved with the \textit{evaluation} module from \gls{SeFEF}. Default parameters were used. Star symbols denote the onset of a seizure, where the ones with lower opacity indicate onsets for which no pre-ictal data was found.}
\label{fig:tscv_example}
\end{figure}

\Gls{TSCV} was implemented with the following default conditions provided by \gls{SeFEF}, as detailed in Appendix \ref{app:tscv} and the results are illustrated in Figure \ref{fig:tscv_example} for the time series dataset (a) and the seizure timestamps dataset (b). Notice how the time series dataset accounts for missing data (i.e., each line represents an existing sample), whereas the synthetic dataset spans the whole period from the first to the last seizure onset. 

\subsection{Model Development}
\label{sec:model_training}

\subsubsection*{Von Mises Estimator: Leveraging Sporadic, Timestamp-Only, Data}

\begin{figure}[!ht]
  \centering
  \includegraphics[width=0.7\linewidth]{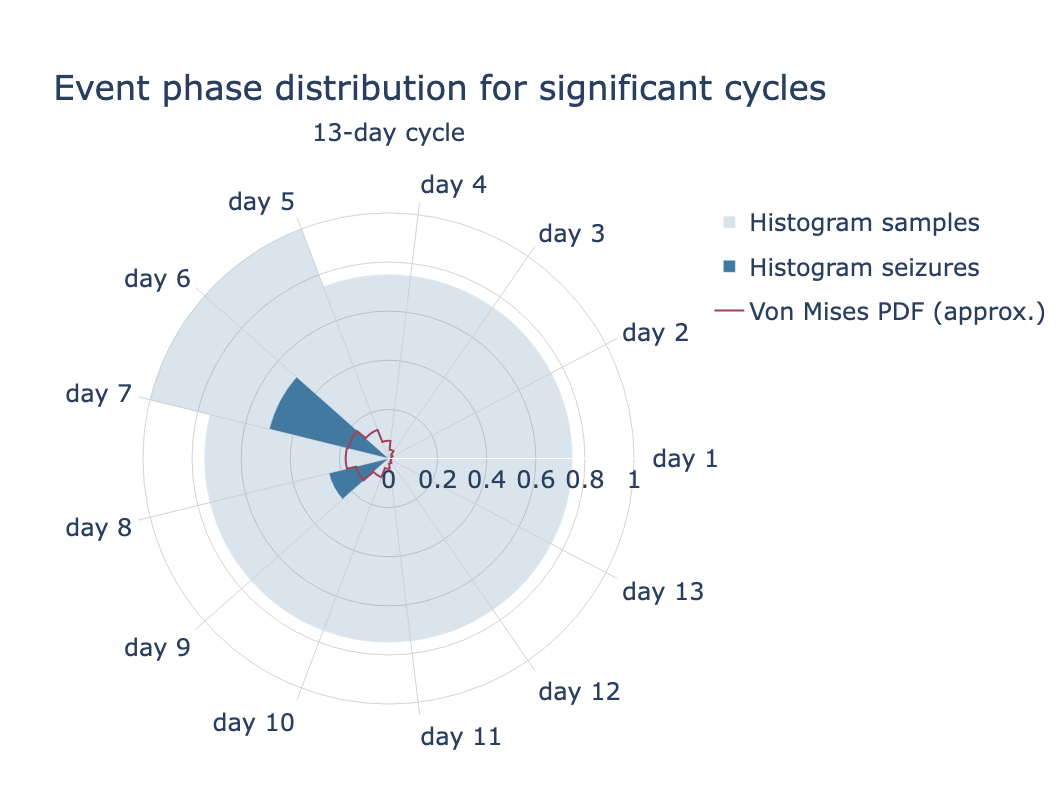}
  \caption{Example \acrshort{PDF} approximation using a von Mises distribution for an 13-day significant cycle (patient 1869, \gls{TSCV} fold 2, 5 seizures in train set). Also shown are frequency histograms for seizure occurrences and the total number of samples, both normalized by the maximum bin count across all bins.}
  \label{fig:vonmises}
\end{figure}

This algorithmic approach is based on the estimation of seizure probability given its phase in a set of significant cycles. Training steps included selecting significant cycles,  fitting a von Mises distribution to the positive samples for each cycle, and estimating its discretized \gls{PDF}, with the cycle unit as the discretization step (exemplified in Figure \ref{fig:vonmises}). 

The von Mises fit provides  $P(\theta_i | S=1)$ for each significant cycle $i$, so the Bayes theorem was used in conjunction with a geometric mean to combine the constructive and destructive contributions of each cycle's information, giving $P(S=1 | \theta_1, ..., \theta_{N_{cycles}})$ as seen in Equations (\ref{eq:vonMises}) and (\ref{eq:vonMises2}).

\begin{gather}
  \label{eq:vonMises}
  factor = \left(\prod_{i=1}^{N_{cycles}} \frac{P(\theta_i | S=1)}{P(\theta_i | S=1) + P(\theta_i | S=0)}\right)^{\frac{1}{N_{cycles}}} \\
  \label{eq:vonMises2}
  P(S=1 | \theta_1, ..., \theta_{N_{cycles}}) = factor \cdot P(S=1)
\end{gather}

\noindent
where $\theta_i, i={1, ..., N_{cycles}}$ is the corresponding phase for the $i$-th significant cycle, and $P(S=1)$ is the seizure prior. Appendix \ref{app:vonMises} provides further details of this implementation.

\subsubsection*{Logistic Regression: Leveraging Time Series Feature Extraction}

This algorithmic approach was trained on 36 features (statistical, Hjorth-based, and event-based, as described in Appendix \ref{app:data}) extracted from time series data. The model was implemented using L2 regularization, optimized with the LBFGS (limited-memory BFGS) solver, and a regularization strength of 1.0. No class balancing strategy was used. 
The \gls{LR} model outputs the probability $P(S=1|X)$, where $X$ represents the feature vector, and is given by:

$$P(S=1|X) = \frac{1}{1+exp(-(w_0 + \sum_{i}^{N}w_i X_i))}$$

\noindent 
where $N$ is the number of features, $w_0$ denotes the bias term (or intercept), and $w_i$ represent the model weights associated with the corresponding features $X_i$.

\subsubsection*{Periodicity-Aware LR: An Ensemble-Based Approach}

This algorithmic approach denotes an ensemble of both previous approaches, as proposed in \cite{Karoly2017CircadianProfile}. A \gls{LR} model was trained in the same form as described above, using the train set. Starting from the standard \gls{LR}, at each time step of the test set, the prior probability $P(S=1)$ was updated based on the estimation of the von Mises estimator. This was done by adjusting the \gls{LR} model's intercept, effectively recalibrating the decision boundary of the model:

$$w_0^\prime = w_0 - log \left(\frac{P(S=1)}{1-P(S=1)}\frac{1-P(S=1)^\prime}{P(S=1)^\prime}\right)$$

\noindent
where $P(S=1)$ is the prior inferred from the training data and $P(S=1)^\prime$ is the updated prior computed using the von Mises estimator for the corresponding time step. More information on this approach is available is Appendix \ref{app:LR}.

\subsubsection*{Threshold Optimization}

To allow for the conversion from a probabilistic model output into a binary classification, a high likelihood threshold was computed for each \gls{TSCV} fold. To optimize the threshold, a \gls{GMM} was employed to model the distribution of predicted probabilities for the train set. Given the tendency of the model to produce low probability scores, a bimodal distribution was assumed, corresponding to low- and high-confidence predictions. The \gls{GMM} was fitted with two components, and the threshold was set at the intersection of the Gaussian distributions. This data-driven approach allowed to define a threshold without requiring additional validation data.

\subsection{Performance Scoring and Visualization}

\begin{figure}[!ht]
  \centering
  \begin{minipage}{0.5\textwidth}
      \centering
      \includegraphics[width=1.\linewidth]{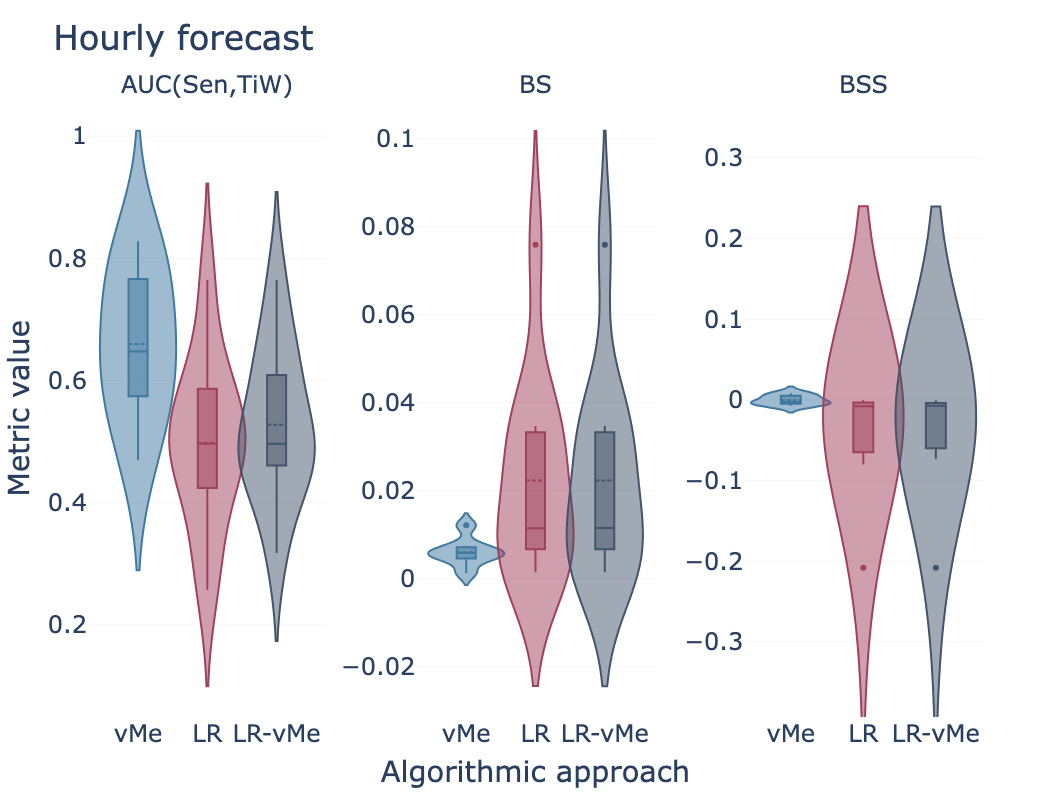}
      (a)
  \end{minipage}%
  \begin{minipage}{0.5\textwidth}
    \centering
    \includegraphics[width=1.\linewidth]{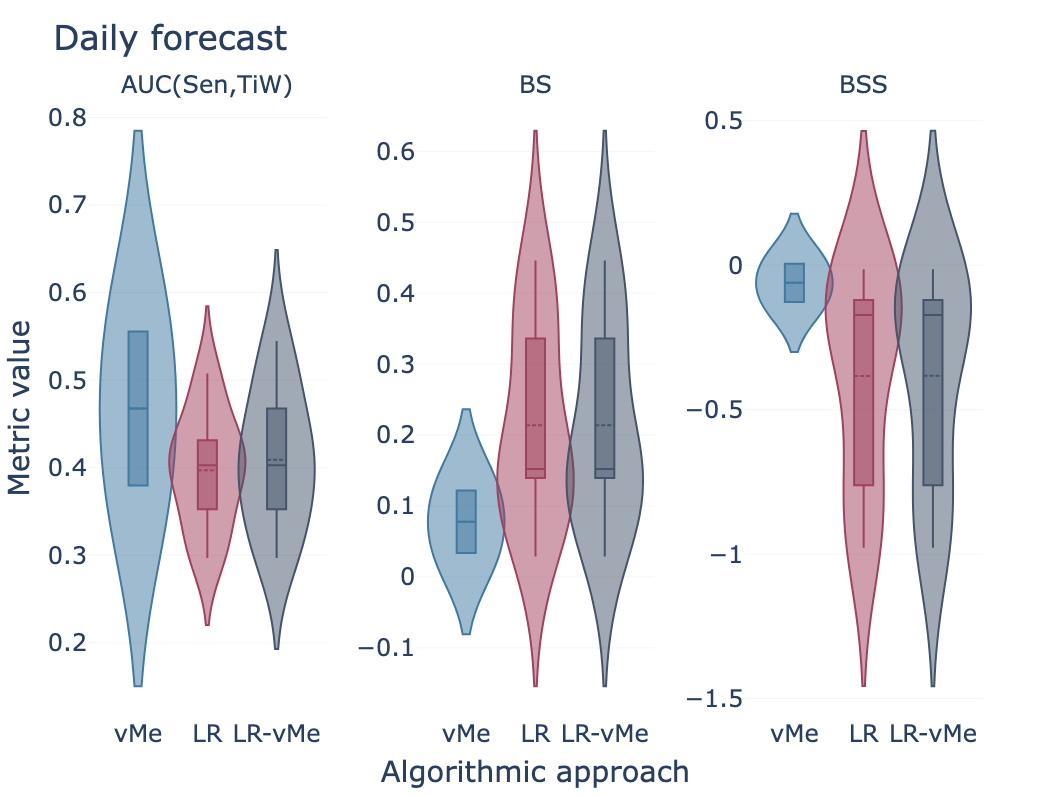}
    (b)
  \end{minipage}
  \caption{Violin plots summarizing the performance scoring achieved by the three algorithmic approaches for (a) hourly forecast and (b) daily forecast. \footnotesize Acronyms: \textbf{\acrshort{LR}:} \acrlong{LR}; \textbf{LR-vMe:} periodicity-aware LR; \textbf{vMe:} von Mises estimator. \textbf{\acrshort{AUCSenTiw}:} \acrlong{AUCSenTiw}; \textbf{\acrshort{BS}:} \acrlong{BS}; \textbf{\acrshort{BSS}:} \acrlong{BSS}.}
  \label{fig:violin}
\end{figure}

Table \ref{tab:results} and Figure \ref{fig:violin} present a summary of performance scoring achieved by the three algorithmic approaches for \gls{AUCSenTiw}, \gls{BS}, and \gls{BSS} (the complete set of performance metrics is also made available in \textit{Supplementary\_Material\_Results.xlsx}). Both hourly (i.e., 1 hour) and daily (i.e., 24 hours) forecasts are presented. 

\begin{table}[!ht]
  \begin{adjustwidth}{-.6in}{-.5in}  
  \centering
  \resizebox{1.2\textwidth}{!}{%
  \begin{threeparttable}
  \caption{Summary of performance scoring achieved by the three algorithmic approaches, given as mean $\pm$ standard deviation. Overall performance across patients is provided in (*). \footnotesize Acronyms: \textbf{\acrshort{LR}:} \acrlong{LR}; \textbf{LR-vMe:} periodicity-aware LR; \textbf{vMe:} von Mises estimator. \textbf{\acrshort{AUCSenTiw}:} \acrlong{AUCSenTiw}; \textbf{\acrshort{BS}:} \acrlong{BS}; \textbf{\acrshort{BSS}:} \acrlong{BSS}. Time series data for patient 1110 did not allow for a \acrshort{TSCV} split (NA in \acrshort{LR}-based models). When no significant cycle was found across any of the \acrshort{TSCV} folds, NA is shown in von Mises estimator. When the standard deviation is not provided, the results correspond to a single fold.}
  \label{tab:results}
  \centering
  \begin{tabular}{cccccccccc}
    \toprule
    \multirow{2}{*}{\textbf{ID}} & \multicolumn{3}{c}{\textbf{AUC(Sen,TiW)}} & \multicolumn{3}{c}{\textbf{BS}} & \multicolumn{3}{c}{\textbf{BSS}} \\ \cmidrule{2-4}\cmidrule{5-7}\cmidrule{8-10}
    & \textbf{vMe} & \textbf{LR} & \textbf{LR-vMe} & \textbf{vMe} & \textbf{LR} & \textbf{LR-vMe} & \textbf{vMe} & \textbf{LR} & \textbf{LR-vMe} \\
    \toprule
    \multicolumn{10}{l}{\textbf{Hourly forecast}} \\
    \midrule
    1110 & 0.61 & NA & NA & 0.00 & NA & NA & -0.00 & NA & NA \\
    1869 & 0.65 & 0.61 $\pm$ 0.00 & 0.61 $\pm$ 0.00 & 0.01 & 0.00 $\pm$ 0.00 & 0.00 $\pm$ 0.00 & 0.00 & -0.01 $\pm$ 0.01 & -0.01 $\pm$ 0.01 \\
    1876 & 0.78 $\pm$ 0.05 & 0.39 $\pm$ 0.19 & 0.49 $\pm$ 0.25 & 0.01 $\pm$ 0.00 & 0.01 $\pm$ 0.00 & 0.01 $\pm$ 0.00 & 0.00 $\pm$ 0.01 & -0.00 $\pm$ 0.01 & -0.01 $\pm$ 0.00 \\
    1904 & NA & 0.76 & 0.76 & NA & 0.00 & 0.00 & NA & -0.02 & -0.02 \\
    1965 & NA & 0.49 $\pm$ 0.02 & 0.49 $\pm$ 0.02 & NA & 0.03 $\pm$ 0.00 & 0.03 $\pm$ 0.00 & NA & -0.05 $\pm$ 0.10 & -0.05 $\pm$ 0.10 \\
    2002 & 0.52 $\pm$ 0.07 & 0.37 $\pm$ 0.05 & 0.43 $\pm$ 0.03 & 0.01 $\pm$ 0.00 & 0.04 $\pm$ 0.05 & 0.04 $\pm$ 0.05 & -0.00 $\pm$ 0.00 & -6.05 $\pm$ 8.44 & -6.04 $\pm$ 8.44 \\
    (*) & 0.66 $\pm$ 0.13 & 0.50 $\pm$ 0.14 & 0.53 $\pm$ 0.13 & 0.01 $\pm$ 0.00 & 0.02 $\pm$ 0.02 & 0.02 $\pm$ 0.02 & -0.00 $\pm$ 0.01 & -1.12 $\pm$ 3.61 & -1.12 $\pm$ 3.61 \\
    \midrule
    \multicolumn{10}{l}{\textbf{Daily forecast}} \\
    \midrule
    1110 & 0.38 & NA & NA & 0.03 & NA & NA & -0.13 & NA & NA \\
    1869 & 0.56 & 0.44 $\pm$ 0.05 & 0.51 $\pm$ 0.05 & 0.12 & 0.10 $\pm$ 0.05 & 0.10 $\pm$ 0.05 & 0.01 & -0.11 $\pm$ 0.07 & -0.11 $\pm$ 0.06 \\
    1876 & NA & 0.47 $\pm$ 0.05 & 0.47 $\pm$ 0.05 & NA & 0.15 $\pm$ 0.01 & 0.15 $\pm$ 0.01 & NA & -0.15 $\pm$ 0.03 & -0.15 $\pm$ 0.03 \\
    1904 & NA & 0.35 & 0.35 & NA & 0.03 & 0.03 & NA & -0.01 & -0.01 \\
    1965 & NA & 0.36 $\pm$ 0.06 & 0.36 $\pm$ 0.06 & NA & 0.38 $\pm$ 0.06 & 0.38 $\pm$ 0.06 & NA & -0.82 $\pm$ 0.16 & -0.82 $\pm$ 0.16 \\
    2002 & NA & 0.38 $\pm$ 0.04 & 0.38 $\pm$ 0.04 & NA & 0.15 $\pm$ 0.01 & 0.15 $\pm$ 0.01 & NA & -0.19 $\pm$ 0.11 & -0.19 $\pm$ 0.11 \\
    (*) & 0.47 $\pm$ 0.12 & 0.40 $\pm$ 0.07 & 0.41 $\pm$ 0.08 & 0.08 $\pm$ 0.06 & 0.21 $\pm$ 0.14 & 0.21 $\pm$ 0.14 & -0.06 $\pm$ 0.09 & -0.38 $\pm$ 0.37 & -0.38 $\pm$ 0.37 \\
    \bottomrule 
  \end{tabular}
\end{threeparttable}}
\end{adjustwidth}
\end{table}

Overall, \Gls{AUCSenTiw} showed an improved score for the von Mises estimator when compared to both other approaches. Although this statement holds true for all patients, there was considerable variability among them, as evidenced by the large standard deviations in overall performance. Both \gls{BS} and \gls{BSS} also reflected an improved performance by the von Mises estimator. 

Despite the slight improvement in \gls{AUCSenTiw} observed in the periodicity-aware \gls{LR} compared to the regular \gls{LR}, no metric exhibited a consistent pattern of performance differences, with no systematic improvement or decline across forecasting approaches.

Despite the overall excellent Brier score (\gls{BS}) in the hourly forecast, an analysis of its two components -- resolution and reliability -- provides some important insights. Specifically, the resolution is generally close to zero, indicating that most forecasts resemble the average seizure probability, which raises concerns about their practical usefulness. Furthermore, the small reliability values observed are likely influenced by the binning approach, including factors such as the number of bins, the distribution of forecasted probabilities across bins, and the limited number of observed events \cite{Dimitriadis2021StableReliability}. This is further supported by the average Brier Skill Scores (\gls{BSS}), all of which were below or very close to zero, indicating no improvement over the naive forecast.

\begin{figure}[!ht]
  \centering
  \begin{minipage}{.51\textwidth}
      \centering
      \includegraphics[width=1\linewidth]{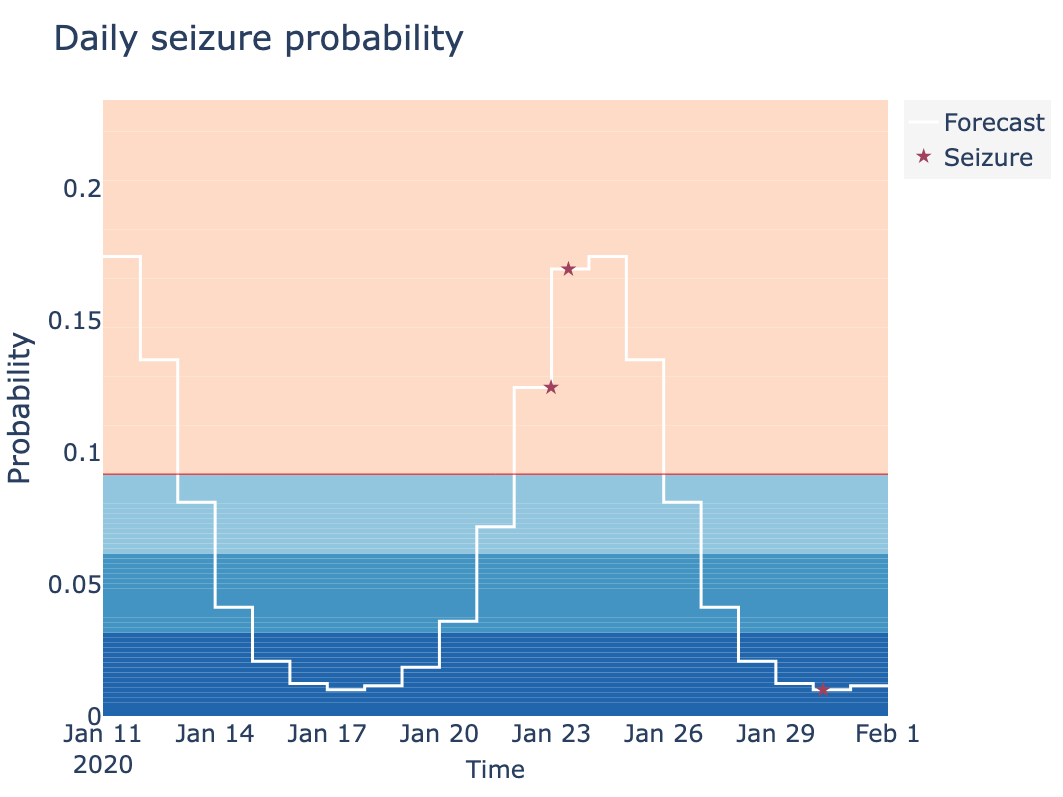}
      (a)
  \end{minipage}%
  \begin{minipage}{.51\textwidth}
    \centering
    \includegraphics[width=1\linewidth]{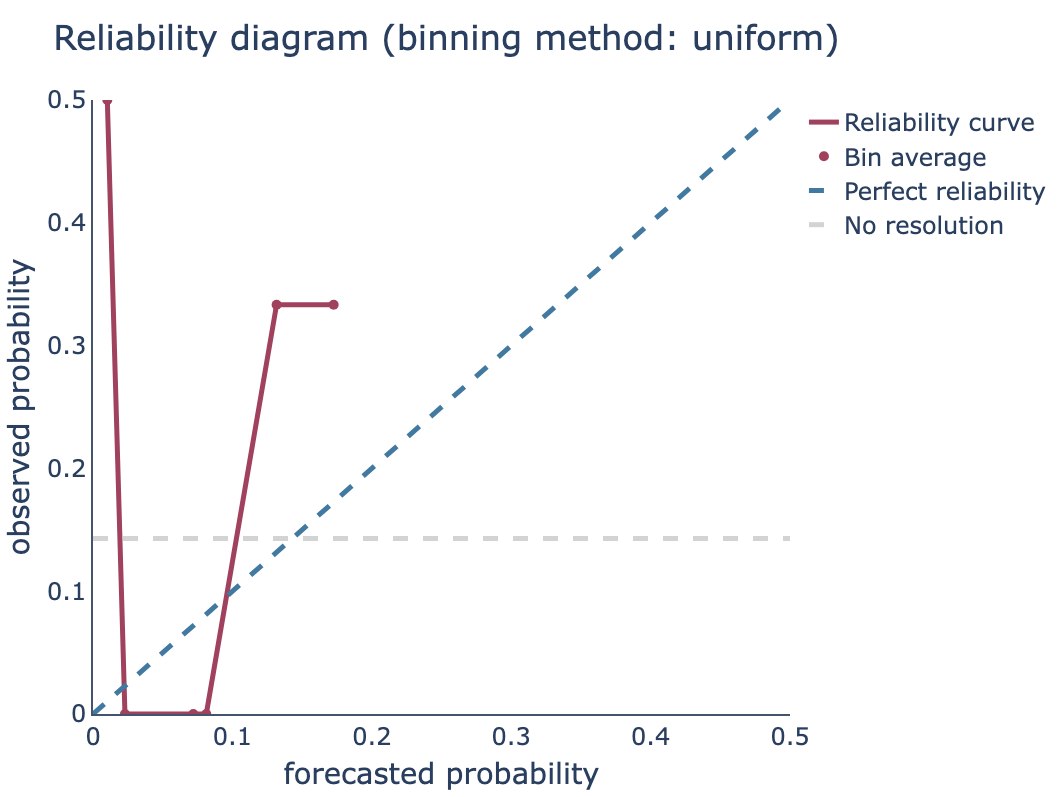}
    (b)
\end{minipage}%
\caption{Example obtained for patient 1869 (by von Mises estimator) with (a) daily forecast and (b) reliability diagram. Only \acrshort{TSCV} fold 1 is shown since no significant forecasts were found for folds 2 and 3. The red horizontal line in the forecast corresponds to the high likelihood threshold computed for that fold. For the reliability diagram, 5 bins were used with uniform distribution of forecasts across the range of forecasted probabilities.}
\label{fig:forecasts}
\end{figure}

Figure \ref{fig:forecasts} (a) shows an example of daily forecasts, generated with \gls{SeFEF}'s \verb|visualization| module. According to the forecast horizon, the probabilities issued at the start of the forecast period are extended across the full period. Figure \ref{fig:forecasts} (b) shows the corresponding reliability (or calibration) diagram, generated with \gls{SeFEF}'s \verb|scoring| module.

\begin{figure}[!ht]
  \centering
  \includegraphics[width=.9\linewidth]{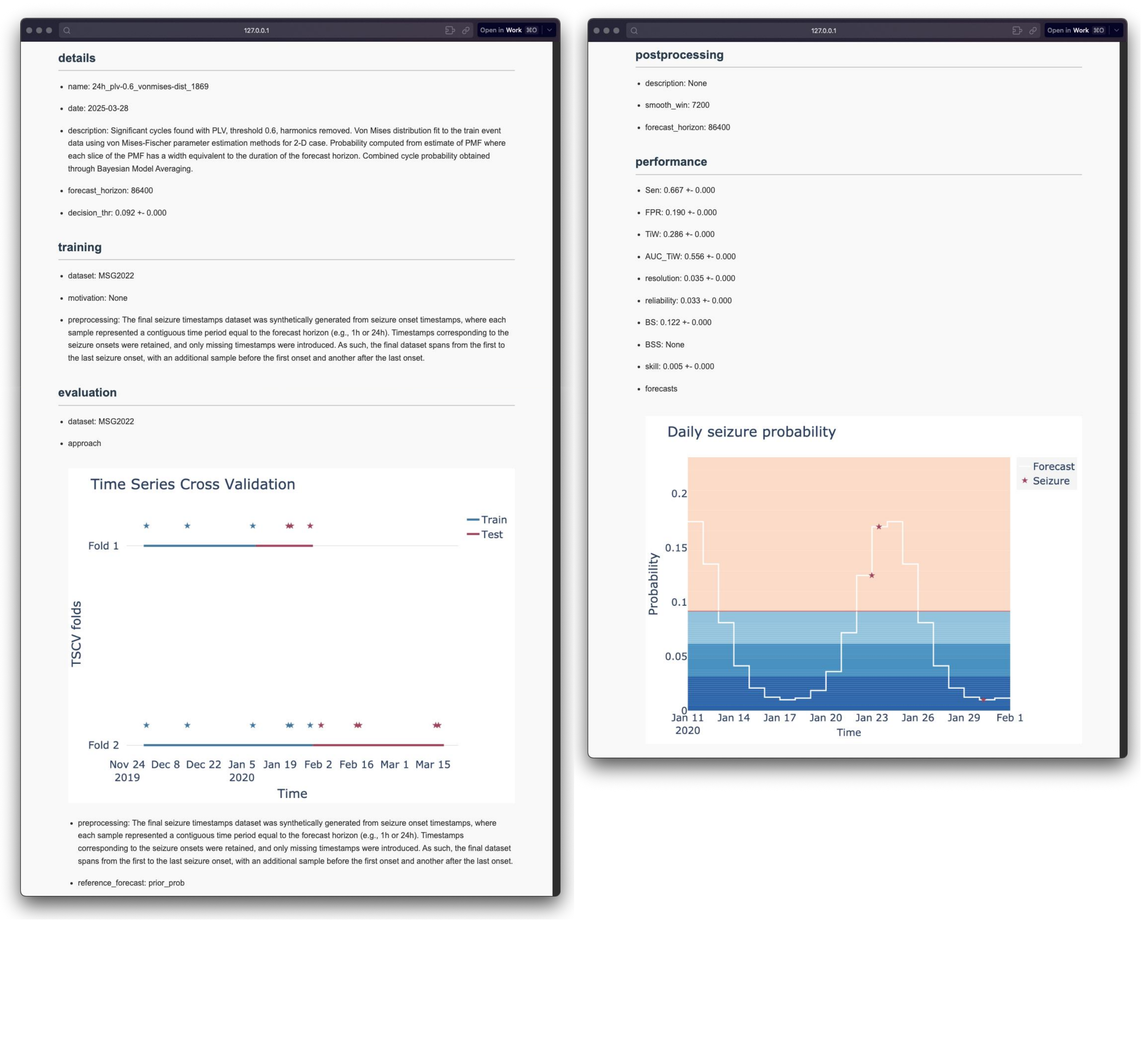}
  \caption{Example model card for periodicity-aware \acrshort{LR}, generated for patient 1869, using \acrshort{SeFEF}.}
  \label{fig:modelcard}
\end{figure}

Finally, Figure \ref{fig:modelcard} shows the several sections that constitute a model card \cite{Mitchell2019ModelCards} generated using \gls{SeFEF}'s \verb|modelcard| module. It includes a unique model name and creation date, along with a description of the implementation, specifying the forecast horizon and relevant methodological choices. It also documents training details, such as the dataset used and preprocessing steps, as well as evaluation procedures, including the \gls{TSCV} strategy and any additional preprocessing applied to the test sets. Performance metrics, reported as mean and standard deviation across \gls{TSCV} folds, provide an overview of model variability, along with a visualization of the forecasts.

\section{Discussion}

This work introduced \gls{SeFEF}, a Python-based framework designed to automate key steps in the development of seizure forecasting algorithms. To demonstrate its versatility, three distinct models were implemented, where each leveraged diverse input data and algorithmic approaches. This proof-of-concept suggests the framework's capacity to standardize and streamline the forecasting pipeline, while accommodating different methodological choices.

\subsubsection*{A Step Towards Comparable Models}

The successful implementation of three diverse models demonstrates the flexibility of the framework. The ability to integrate algorithmic approaches, ranging from phase modeling and classical machine learning to ensemble methods, suggests that the framework can be adapted to a wide range of seizure forecasting methodologies. Additionally, the framework supports both daily and hourly forecasts, which broadens its applicability across different prediction timescales.

Although the primary objective of this study was not to optimize algorithm performance, the results provide insights into how different approaches can be compared with the help of the framework. By automating key steps in seizure forecasting — such as data labeling, cross-validation splitting, post-processing of forecasts, and performance metric computation — \gls{SeFEF} minimizes the risk of unintentional methodological variations that could compromise comparability between approaches and ensures a more consistent and fair comparison between models. For instance, the improved performance in \gls{AUCSenTiw} of the von Mises estimator compared to both other models suggests that the periodicity in seizure occurrence could offer valuable complementary information to traditional time series-based feature analysis. Therefore, the limited improvement observed between the cycle-aware \gls{LR} and the traditional approach may suggest exploring alternative strategies for the ensemble of low probability densities and \gls{ML} models.

The results on reliability and resolution also highlight the importance of how we use probabilistic metrics to inform model development. While the \gls{BS} is often taken at face value, understanding its components (e.g., through the calibration diagram) and considering the specifics of the data (e.g., the number of true events and the distribution of forecasts) is essential for accurately interpreting model performance.

Naturally, this initial validation has inherent limitations, as it is based on a single-user experiment without statistical comparisons or evaluations across multiple independent users. At this stage, our objective is to make the framework publicly available to foster community engagement, facilitate experimentation, and gather feedback. To this end, we have made the project openly accessible on GitHub, where technical discussions are encouraged through the \textit{Discussions} section, and we also provide a Google Form to collect broader feedback from the community\footnote{GitHub Discussions available at \url{https://github.com/anascacais/SeFEF/discussions}; Google Form available at \url{https://forms.gle/AiQbwyYq37nhXjrF8}.}. In the long term, we aim to contribute to the establishment of a consensus on a standardized methodology for the development and validation of seizure forecasting algorithms in people with epilepsy.

\subsubsection*{A Note on Interpretability}

When using probabilistic models designed to approximate true probabilities, it is not possible to disregard how users will interpret the results. Given that these models account for factors such as data uncertainty, seizure frequency, and intrinsic variability, they are unlikely to produce probability values close to 1. How, then, will users interpret such probabilities?

A common assumption is that most people struggle to understand probabilistic information, which could hinder their ability to make appropriate risk-based decisions \cite{Mylne2023CommunicatingProbability}. However, literature reviews, such as that by Mylne \cite{Mylne2023CommunicatingProbability}, reveal a strong consensus that this is not necessarily the case—provided the information is communicated effectively. Language is a crucial factor, and providing a worded description as been shown to decrease misinterpretation \cite{Mylne2023CommunicatingProbability}\footnote{Consider a well-calibrated model where the estimated probabilities range from 0.0 to 0.2. In this case, interpreting a probability of 0.2 could be useful for users by associating it with a more tangible concept, such as: for every 5 days under similar conditions, a seizure is expected to occur on one of them. This type of worded description alongside the probability value may provide users with a clearer understanding of the risk, making the information more accessible and actionable.}.

Particularly in epilepsy, Chiang et al. \cite{Chiang2021EvaluationRecommendations} surveyed people with epilepsy, caregivers, and healthcare providers in regard to functionality, appropriateness, and usability of several visualization methodologies. Data visualization solutions included daily heat maps, line and rose plots, risk gauge, radar charts, etc. Besides identifying hourly radar charts as the option least prone to interpretation error, this study also showed that accuracy of interpretation was dependent on seizure frequency, and that usefulness of hourly/daily visuals was dependent on the presence/absence of seizure cycles. Therefore, studies like this one can also provide valuable insights into how visual representations of forecasted probabilities influence user interpretation.


\section*{Author Contributions}

\textbf{Ana Sofia Carmo:} conceptualization (lead); investigation (lead); formal analysis (lead); visualization (lead); writing - original draft (lead); writing - review and editing (equal). \textbf{Lourenço Abrunhosa Rodrigues:} conceptualization (supporting); formal analysis (supporting); writing - original draft (supporting); writing - review and editing (equal). \textbf{Ana Rita Peralta:} conceptualization (supporting); supervision (supporting); writing - review and editing (equal). \textbf{Ana Fred:} supervision (supporting); writing - review and editing (supporting). \textbf{Carla Bentes:} conceptualization (supporting); supervision (supporting); writing - review and editing (equal). \textbf{Hugo Plácido da Silva:} conceptualization (supporting); supervision (lead); writing - review and editing (equal).

\section*{Acknowledgments}
This work was supported by the \acrfull{FCT}, Portugal, under the grant 2022.12369.BD, by FCT/MCTES through national funds, partially funded by the Deutsche Forschungsgemeinschaft (DFG, German Research Foundation) - 459360854 as part of the Research Unit "Lifespan AI: From Longitudinal Data to Lifespan Inference in Health" (DFG FOR 5347), University of Bremen (http://lifespanai.de), and when applicable, co-funded by EU funds under the project UIDB/50008/2020 (DOI identifier https://doi.org/10.54499/UIDB/50008/2020). It was done under the project "PreEpiSeizures" in a collaboration between \acrfull{IT} and the Neurophysiology Monitoring Unit-EEG/Sleep Laboratory from \acrfull{ULSSM}.

\subsubsection*{Competing Interests}
The authors have no conflicts of interest to disclose.

\subsubsection*{Ethics Statement}
This research did not involve any experiments on human or animal subjects. Therefore, ethical approval was not required. No interaction with human participants took place.

\subsubsection*{Declaration of Generative AI and AI-Assisted Technologies in the Writing Process}

During the preparation of this work the authors used ChatGPT (OpenAI) in order to improve the readability of the manuscript. After using this tool, the authors reviewed and edited the content as needed and take full responsibility for the content of the published article. 

\clearpage
\newpage
\bibliographystyle{ieeetr}
\bibliography{references}

\clearpage
\newpage
\appendix
\section{Performance Scoring}
\label{app:performance}

Deterministic predictions of whether an event will happen within a specific time frame (i.e., a categorical yes/no) are incredibly useful when they are timely and their accuracy is close to perfect, as they allow for prompt and informed decision-making \cite{Proix2021ForecastingSeizure}.  However, in the context of seizure occurrence, this is hardly conceivable \cite{Proix2021ForecastingSeizure}. Instead, a probabilistic forecast strives to provide an aggregated measure of seizure prediction with its uncertainty \cite{Mason2015GuidanceVerification}, which allows users to interpret the prediction and act according to their aversion/comfort with uncertainty.

Probabilistic forecasts have been historically evaluated through deterministic metrics \cite{Carmo2024AutomatedAlgorithms}, which require the conversion of the predicted probability into high/low likelihood states, and often relies on threshold optimization (high/low likelihood thresholds). The categorical likelihood states are leveraged to define the classical concepts of \glspl{TP}, \glspl{FP}, and \glspl{FN}, which can then be used to compute standard performance metrics such as \gls{Sen} and \gls{FPR}. 

Despite the commonality of deterministic metrics, some authors have started evaluating their proposed approaches through probabilistic measures of performance \cite{Carmo2024AutomatedAlgorithms}, borrowed from the domain knowledge of weather forecasting. As proposed by Mason \cite{Mason2015GuidanceVerification}, the quality of a forecast is complex and multidimensional. It includes attributes such as how strongly the outcome is conditioned on the forecast (i.e., resolution), if the confidence of the forecast is in agreement with the observed rate of the corresponding outcome (i.e., reliability), and how much better the forecast is when compared to a reference (i.e., skill). Given this duality of deterministic and probabilistic performance evaluation, and also reflecting the trends in the literature, \gls{SeFEF} provides a set of both deterministic and probabilistic scoring metrics.

\subsubsection*{Deterministic Metrics}

Unlike in the classical sample-based scoring where \glspl{TP}, \glspl{FP}, and \glspl{FN} are computed by comparing high/low likelihood states to the pre-ictal vs inter-ictal annotations\footnote{In sample-based scoring, scoring is heavily reliant on the pre-ictal duration, since each sample's forecast is compared against its annotation (pre-ictal vs inter-ictal).}, in event-based scoring these are computed according to the occurrence (or not) of a seizure event within the forecast horizon \cite{Dan2024SzCORESeizurea,Ren2022PerformanceEvaluation}. The high/low likelihood thresholds should be obtained by the user, for instance through optimization during model training. Otherwise, a naive threshold of 0.5 is applied. 

\gls{SeFEF} offers the following deterministic metrics: \acrfull{Sen}, \acrfull{FPR}, \gls{TiW}, and \gls{AUC}. \Gls{Sen} provides a measure of the model's ability to correctly identify pre-ictal periods. In practice, this corresponds to the proportion of seizures with forecasts of high likelihood and is given by: $$\textnormal{Sen} = \frac{\textnormal{TP}}{\textnormal{TP} + \textnormal{FN}}$$

\noindent
where \gls{TP} corresponds to the number of forecasts of high likelihood that were followed by seizure events (within the corresponding forecast horizon) and \gls{FN} provides the number of seizures that were not preceded by a forecast of high likelihood. 

\gls{FPR} indicates the proportion of time that the user incorrectly spends in alert and is given by: $$\textnormal{FPR} = \frac{FP}{n}$$ 

\noindent 
where $n$ is the total number of forecasts and \gls{FP} corresponds to the number of high likelihood forecasts that were not followed by seizure events. Alternatively, \gls{TiW} aims to characterize the proportion of time that the user spends in alert, i.e., in a high likelihood state, independently of the \textit{"goodness"} of the forecast. It is given by: $$\textnormal{TiW} = \frac{\textnormal{TP} + \textnormal{FP}}{n}$$

Finally, the \gls{AUC} is provided as a measure of the trade-off between \gls{Sen} and \gls{TiW}, abstracting the need for threshold optimization. Instead, each unique forecast value is iteratively used as a high likelihood threshold and both \gls{Sen} and \gls{TiW} are recomputed. \gls{AUC} is then computed as the numerical integration of \gls{Sen} vs \gls{TiW} using the trapezoidal rule\footnote{For implementation of the trapezoidal rule for computation of the \acrlong{AUC} from \acrlong{Sen} and \acrlong{TiW}, \texttt{scikit-learn}'s \texttt{auc} function was used.}.

As proposed by Dan et al. \cite{Dan2024SzCORESeizurea}, performance metrics based on the count of \glspl{TN} were intentionally excluded, as the definition of this concept is ambiguous in the context of event-based scoring.

\subsubsection*{Probabilistic Metrics}

Measures of resolution, reliability, \gls{BS}, and \gls{BSS} are made available in \gls{SeFEF} and were implemented as proposed by Proix et al. \cite{Proix2021ForecastingSeizure}, Mason \cite{Mason2015GuidanceVerification}, and Stephenson et al. \cite{Stephenson2008TwoExtra}. On one hand, resolution is defined as the ability of the forecast to differentiate between individual observed probabilities and the average observed probability, and is given by: $$\textnormal{resolution} = \frac{1}{n}\sum_{k=1}^{b}n_k(\bar{y_k}-\bar{y})^2$$

\noindent
where $n$ is the total number of forecasts, $b$ is the total number of probability bins (of equal number of observations), $n_k$ is the number of forecasts for the $k^\textnormal{th}$ probability bin, $\bar{y_k}$ is the number of times an event occurred in the $k^\textnormal{th}$ probability bin divided by the number of times that probability range was forecast, and $\bar{y}$ is the observed relative frequency of true events for all forecasts. More specifically, given the set of forecasts whose output probability falls within the $k^\textnormal{th}$ probability bin, $\bar{y_k}$ is computed as the ratio between the number of seizures that actually occurred in the time period corresponding to those forecasts and the total number of forecasts in that bin.

\begin{figure}[!ht]
  \includegraphics[width=1.\linewidth]{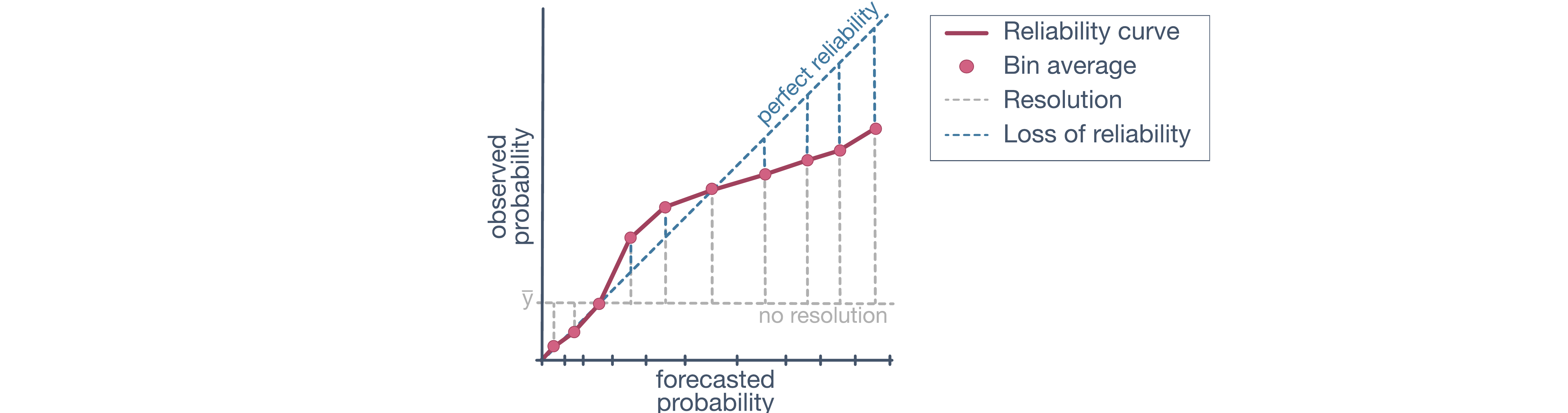}
  \caption{Illustration of a reliability diagram. The forecasts were binned into probability bins of equal number of observations. The threshold of no resolution ($\bar{y}$) corresponds to a forecast equal to the observed frequency of seizures. Forecasts along the dashed diagonal line correspond to perfectly reliable (or calibrated) forecasts, while forecasts above and below that line correspond to under-confident and over-confident predictions, respectively. This illustration is inspired by the one in \cite{Proix2021ForecastingSeizure}.}
  \label{fig:resolution_reliability}
\end{figure}

On the other hand, reliability (also known as calibration) measures the agreement between forecasted and observed probabilities, which is given by: $$\textnormal{reliability (or calibration)} = \frac{1}{n}\sum_{k=1}^{b}n_k(\bar{p_k}-\bar{y_k})^2$$

\noindent
where $\bar{p_k}$ is the average probability of the forecasts within the $k^\textnormal{th}$ probability bin. The smaller this score is (with the minimum of 0 and maximum of 1), the better the performance. Graphically, these measures can be represented as illustrated in Figure \ref{fig:resolution_reliability}.

The \gls{BS} is a commonly used probabilistic metric which provides a measure of the accuracy of the forecast and is given by: $$\textnormal{BS} = \frac{1}{n}\sum_{k=1}^{b}\sum_{i=1}^{n_k}(p_{ki} - y_{ki})^2$$

\noindent
where $y_{ki}$ corresponds to the outcome of forecast $i$ in bin $k$ and can take the values of either 0 or 1, according to the occurrence of a seizure. In turn, it can be decomposed into 5 terms \cite{Stephenson2008TwoExtra}: $$\textnormal{BS} = \textnormal{reliability} - \textnormal{resolution} + \textnormal{uncertainty} + \textnormal{WBV} - \textnormal{WBC}$$

\noindent
where the last three terms correspond, respectively, to uncertainty, \gls{WBV}, and \gls{WBC}, and are given by: $$\textnormal{uncertainty} = \bar{y}(1-\bar{y}) ~;~ \textnormal{WBV} = \frac{1}{n}\sum_{k=1}^{b}\sum_{i=1}^{n_k}(p_{ki}-\bar{p_k})^2 ~;~ \textnormal{WBC} = \frac{2}{n}\sum_{k=1}^{b}\sum_{i=1}^{n_k}(y_{ki}-\bar{y_k})(p_{ki}-\bar{p_k})$$

Given that \gls{WBV} and \gls{WBC} account for the binning strategy, both terms vanish if each bin contains a single forecast value. Finally, since skill is a relative metric, it is always computed in reference to another forecast. In \gls{SeFEF}, we propose the use of \gls{BSS} against a naive forecast (also known as "climatology" reference in the domain of weather forecasts), which perpetually predicts the relative frequency of seizures. This is an appealing reference strategy since it is equivalent to solely relying on the historical seizure frequency to issue a forecast \cite{Mason2004UsingClimatology}. The \gls{BSS} is given by: $$\textnormal{BSS (or reliability skill score)} = 1 - \frac{BS}{BS_{\textnormal{ref}}} = 1 - \frac{BS}{\bar{y}(1-\bar{y})}$$

\noindent
where the reliability and resolution terms of the naive forecast disappear due to $\bar{p_k}$, $\bar{y_k}$, and $\bar{y}$ all taking the same value. This score ranges from -$\infty$ to 1, where negative values indicate that the forecast is less skilled than the reference and 0 and 1 correspond to no-skill and perfect skill compared to reference, respectively.

\section{Data Preparation \& Labeling}
\label{app:data}
\subsubsection*{Time Series Dataset}

Time series data from 4 of the available channels (i.e., \gls*{ACC} magnitude, \gls*{BVP}, \gls*{EDA}, and \gls*{HR}) were segmented into 60-second windows and preprocessed as follows, using the Python package \verb|BioSPPy|: \gls{BVP} and \gls{EDA} were filtered using a 4th-order Butterworth filter with modality-specific frequency range (Table \ref{tab:filters}), followed, in the case of \gls{EDA}, by a moving-average filter with duration of 750 ms. The remaining modalities did not undergo any preprocessing step. Windows with any missing data from any of the included modalities were excluded. 

\begin{table}[!ht]
  \centering
  \caption{Frequency range, in Hz, of 4th-order Butterworth filters applied to each modality as a preprocessing step. \footnotesize Acronyms: \textbf{\acrshort{ACC}}: \acrlong{ACC}; \textbf{\acrshort{BVP}}: \acrlong{BVP}; \textbf{\acrshort{EDA}}: \acrlong{EDA}; \textbf{\acrshort{HR}}: \acrlong{HR}.}
  \label{tab:filters}
  \begin{tabular}{cccc}
    \toprule
    \textbf{\acrshort{ACC} magnitude} & \textbf{\acrshort{BVP}} & \textbf{\acrshort{EDA}} & \textbf{\acrshort{HR}} \\
    None & [1 - 7] & [0, 5] & None \\
    \bottomrule
  \end{tabular}
\end{table}

Statistical features were extracted from all windows and all modalities, including mean, signal power, standard deviation, kurtosis, skewness, mean of first-order differences, and mean of second-order differences. Additionally, from the \gls{EDA}, Hjorth-based features were extracted (i.e., activity, mobility and complexity), as well as event-based features (i.e., number of peaks, mean amplitude, mean rise time, sum of amplitudes, sum of rise times). This yielded a total of 36 features per 60-second window, as summarized in Table \ref{tab:features}. The final time series dataset consisted on the complete set of 60-second windows (i.e., the samples) extracted from the original data files, each represented by a feature vector, the Unix-timestamp corresponding to the start time of the window, and its label. 

\begin{table}[!ht]
  \centering
  \caption{Features extracted, presented by category and modality, totaling 36 features. \footnotesize Acronyms: \textbf{\acrshort{ACC}}: \acrlong{ACC}; \textbf{\acrshort{BVP}}: \acrlong{BVP}; \textbf{\acrshort{EDA}}: \acrlong{EDA}; \textbf{\acrshort{HR}}: \acrlong{HR}.}
  \label{tab:features}
  \begin{tabular}{cC{8cm}c}
    \toprule
    \textbf{Category} & \textbf{Features} & \textbf{Modalites} \\
    \midrule
    Statistical & mean, signal power, standard deviation, kurtosis, skewness, mean of first-order differences, mean of second-order differences & \acrshort{ACC}, \acrshort{BVP}, \acrshort{EDA}, \acrshort{HR} \\ 
    \midrule
    Hjorth-based & activity, mobility, complexity & \acrshort{EDA} \\
    \midrule
    Event-based & number of peaks, mean amplitude, mean rise time, sum of amplitudes, sum of rise times & \acrshort{EDA} \\
    \bottomrule
  \end{tabular}
\end{table}

Given the probabilistic framework of seizure forecasting, contrarily to seizure prediction, the output is an equally-spaced and continuous measure of likelihood \cite{Carmo2024AutomatedAlgorithms}. However, binary labeling is still typically used during model training (inter-ictal/pre-ictal), in which the pre-ictal duration and onset setback (i.e., the time before the onset of a seizure) are determined according to empirical or clinical knowledge. This concept, analogous to seizure prediction, aims to train the model to recognize the patterns and dynamics that may be associated to the occurrence of a seizure. 

\begin{figure}[!ht]
  \centering
  \includegraphics[width=.8\linewidth]{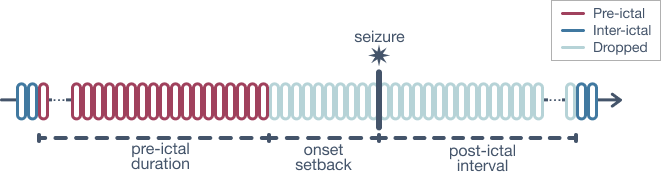}
  \caption{Illustration of labeling procedure when dealing with samples extracted from time series. The \textit{Dropped} samples correspond to samples that are removed from the train set only.}
  \label{fig:labeling}
\end{figure}

Therefore, each individual sample is labeled according to the chosen \textit{pre-ictal duration} (60 minutes) and \textit{onset setback} (10 minutes), as shown in Figure \ref{fig:labeling}.

\subsubsection*{Seizure Timestamps Dataset}

The final seizure timestamps dataset was synthetically generated from seizure onset timestamps, where each sample represented a contiguous time period equal to the forecast horizon (e.g., 1h or 24h). Timestamps corresponding to the seizure onsets were retained, and only missing timestamps were introduced. As such, the final dataset spans from the first to the last seizure onset, with an additional sample before the first onset and another after the last onset.

Take as an example the following set of Unix timestamps corresponding to seizure onsets:
\begin{gather*}
  [1575243000, 1576458600, ..., 1584441600]\\ 
  \Leftrightarrow [\textnormal{2019-12-01 23:30:00, 2019-12-16 01:10:00, ..., 2020-03-17 10:40:00}]
\end{gather*}

Each sample in the seizure timestamps dataset is represented by a feature vector, the Unix-timestamp corresponding to the start time of the window, and its label, as illustrated in Table \ref{tab:timestamps_dataset}. For a forecast horizon of 24 hours, timestamps cover the period from the day preceding the first seizure onset to the day following the last onset. Each sample begins at the start of a day and spans a full 24-hour period (\verb|24h=86400| seconds). The feature vector comprises a single entry per sample, corresponding to the entire timestamp, where the hour information reflects either the actual time of seizure onsets or is synthetically generated for non-seizure periods.

Traditional time series-based approaches typically focus on identifying patterns within the pre-ictal interval to predict seizure onset. In contrast, periodicity-based approaches aim to uncover recurrent temporal patterns in seizure occurrences. Accordingly, in the dataset labeling process, a \textit{no prediction latency} assumption is applied: samples are labeled as 1 if a seizure occurs within the given day and 0 otherwise.

\begin{table}[!ht]
  \centering
  \caption{Both the \textit{timestamps} and \textit{feature\_vector} columns were converted into datetime objects for visualization purposes. First onset is at 2019-12-01 23:30:00 and last onset at 2020-03-17 10:40:00.}
  \label{tab:timestamps_dataset}
  \begin{tabular}{ccc}
    \toprule
    \verb|timestamps| & \verb|feature_vector| & \verb|label| \\
    2019-11-30 & 2019-11-30 05:00:00 & 0 \\
    2019-12-01 & 2019-12-01 23:30:00 & 1 \\
    ... & ...  & ... \\
    2020-03-18 & 2020-03-18 02:00:00 & 0\\
    \bottomrule
  \end{tabular}
\end{table}

\section{Implementation of Time Series Cross-Validation}
\label{app:tscv}

Beyond maintaining temporal integrity, when applying \gls{TSCV} in the context of seizure forecasting, a set of requirements should be considered for the initial and following splits. These may be determined by the user, but our framework also includes pre-set conditions:

\begin{itemize}[noitemsep]
  \item \textbf{Minimum number of lead seizures in training:} A minimum number of events should be assured for training purposes. This value defaults to 3 \cite{Viana2023SeizureForecasting}.
  \item \textbf{Initial train duration:} The duration of the initial training is set to $1/3$ of the recorded duration \cite{Viana2023SeizureForecasting}, and the cutoff is shifted if necessary to meet the previous requirement. However, if the envisioned methodology of analysis implies the extraction of cycles (e.g., from seizure time), the maximum cycle period should be conditioned on this initial train duration. For reference, Karoly et al. \cite{Karoly2020ForecastingCycles} and Xiong et al. \cite{Xiong2023ForecastingSeizure} propose a maximum of one-fifth and one-half of the recording duration, respectively. 
  \item \textbf{Minimum number of lead seizures in testing:} A minimum number of events should be assured for testing purposes in each fold. This value defaults to 3 in order to have more than a single event in the reliability diagram.
  \item \textbf{Test duration:} The duration of the test sets defaults to $1/2$ of the initial train duration, and the cutoff is shifted if necessary to meet the previous requirement. Since this variable is constant across \gls{CV} folds, the last fold is determined by the last available segment for testing with the specified duration\footnote{If preferred by the user, the default behavior can be changed to extend the test duration of the last fold to include all available data.}. 
\end{itemize}

\noindent
where a lead seizure is defined as an event that is not preceded by another seizure onset within a 4-hour timeframe \cite{Viana2023SeizureForecasting}.

\section{Derivation of the Von Mises Estimator}
\label{app:vonMises}
\subsubsection*{Parameter Estimation}
\label{vonMises:parameter_est}

For each significant cycle, a von Mises distribution was fit to the positive samples, whose \gls{PDF} for a phase $\theta$ is given by:

$$f(\theta | \mu, k) = \frac{exp(k cos(\theta - \mu))}{2\pi I_0(k)}$$

\noindent
where $\mu$ and $k$ are the distribution parameters, analogous to $\mu$ and $1/\sigma^2$ from the normal distribution; and $I_0(k)$ is the modified Bessel function of the first kind of order 0. $I_0(k)$ works as a scaling constant so that the distribution sums to unity, given that $\int_{-\pi}^{\pi} exp(k cos(\theta - \mu))d\theta = 2\pi I_0(k)$.

A likelihood maximization approach was used to estimate the $\mu$ and $k$ parameters. This corresponds to finding the maximization solution:

$$\mathop{\mathrm{argmax}}_{\mu, k} \,\, \sum_i \left[\left(k\cdot cos\left(\theta_i - \mu\right)\right) - log\left(I_0(k)\right) - log\left(2\pi\right)\right]$$

While this approach leads to an estimation of $\mu$ based on the angle of a mean complex phasor $\mu = arctan(\Im\{\sum_i e^{j \theta_i}\}/\Re\{\sum_i e^{j \theta_i}\})$, the direct computation of $k$ depends on the ratio of modified Bessel functions of order 1 and 0, $I_1(k)/I_0(k)$, so an approximation adapted from \cite{Sra2012ShortNote} was used. Because the Newton's method approximation used in \cite{Sra2012ShortNote} is tailored for high dimensionality, the number of iterations was increased to guarantee enough precision in this two-dimensional case.

Furthermore, because the concentration parameter $k$ of the von Mises distribution is estimated without accounting for sample size, and therefore does not capture the uncertainty inherent in a finite dataset, a correction term inspired by \cite{Ng2023PenalizedMaximum} was introduced. The term chosen was $1/N$, $N$ being the number of samples available for the parameter estimation. This correction was also applied to the estimation of the significance level of candidate cycles, to penalize seemingly significant cycles with little evidence to support them.

With this in consideration, Algorithms \ref{alg:estimate_rho} and \ref{alg:estimate_k} present the estimation of the relevant parameters.

\begin{algorithm}
\caption{Obtaining $mu$ and $\rho$ from phase values and with correction from \cite{Ng2023PenalizedMaximum}}\label{alg:estimate_rho}
\begin{algorithmic}
  \Require $\Theta \gets \{\theta_0, \theta_1, ..., \theta_N\}, \theta_i \in [-\pi, \pi[$ \Comment{Events as phases in radians}
  \State $N \gets length(\Theta)$
  \State $\overline{Z} \gets \frac{1}{N}\sum_{i=1}^{N}e^{j\cdot \theta_i}$
  \State $\rho \gets ||\overline{Z}|| - \frac{1}{N}$
  \State $\mu \gets angle(\overline{Z})$
  \Ensure $\rho$, $\mu$
\end{algorithmic}
\end{algorithm}

This corrected value of $\rho$ was used both to evaluate the significance level of each candidate cycle, and as a basis for the estimation of the parameter $k$.

\begin{algorithm}
  \caption{Estimating $k$ from $\rho$ using approximation from \cite{Sra2012ShortNote}}\label{alg:estimate_k}
  \begin{algorithmic}
    \Require $\rho$
    \Statex
    \Function{$A_2$}{$x$}
      \State{$a \gets \frac{I_1(x)}{I_0(k)}$} \Comment{$I_n(x)$: modified Bessel function of first kind of order $n$}
      \State \Return{$a$}
    \EndFunction

    \If{$\rho \approx 1$}
      \State $k \gets 10^{12}$\Comment{Avoid unlimited concentration factor}
    \ElsIf{$\rho \leq 0$}
      \State $k \gets 10^{-12}$\Comment{Avoid $k$ being exactly zero}
    \EndIf
    \State $k \gets \frac{\rho(1-\rho^2)}{1-\rho}$
    \While{$i < n\_iter$}
      \State $k \gets k - \frac{A_2(k) - Rho}{1 - A_2^2(k) - \frac{A_2(k)}{k}}$
    \EndWhile
    \Ensure $k$
  \end{algorithmic}
  \end{algorithm}

The same optimization was done for both the positive and negative samples, achieving an approximation of $P(\theta|S=1)$ and $P(\theta|S=0)$, respectively. During training, an additional parameter was computed for both distributions, namely the priors, i.e., $P(S=1)$  and $P(S=0)$\footnote{The priors were computed from the labels, i.e., they are not sensitive to multiple onsets within the same sample.}.

\subsubsection*{Selection of Significant Cycles}

Choosing which cycle lengths to consider for this analysis must be done with care. First, not all possible cycle lengths will be informative enough to warrant the effort of fitting a von Mises distribution, and second, different forms of coupling between cycle lengths would result in a level of dependence between cycles that would make any further inference a lot more difficult.

The first issue is easily solved by setting a threshold over which we consider a cycle to be significant enough. In this work we associate the threshold to the $\rho$ parameter described above, and set it to 0.6, which corresponds to a minimum of 3 events perfectly aligned within the cycle, or more seizures positively contributing to it.

The second issue is solved by equating the $\rho$ parameter for each cycle duration to a frequency decomposition of the sequence of events. Note that $\rho$ is the magnitude of the average complex phasor considering the events' location as phases in a specific cycle length. And note also the obvious (inverse) relation between cycle lengths and event frequency. With this in mind, the local maxima of the distribution of $\rho$ over cycle lengths was considered to correspond to "fundamental cycle lengths", which need to be composed to reconstruct the overall distribution of events. Furthermore, harmonics were removed from the set of fundamental lengths, this way ensuring as much as possible the independence of the information included in each of the von Mises distributions that result from this process.

The cycle selection process follows the following steps: estimation of $\rho$ for all candidates; identification of local maxima; removal of harmonics; and selection of cycles that pass the threshold.

\subsubsection*{Combining Results from Multiple Cycles}

The estimation of seizure probability at the forecast horizon was based on two sets of information: a prior, and the additional information provided by the von Mises distributions obtained for each significant cycle.

The prior was estimated as the probability of observing at least one seizure at any given interval with the duration of the forecast horizon. It was computed as the number of intervals where at least a seizure was observed divided by the total number of observed intervals.

As noted in \ref{vonMises:parameter_est}, the von Mises distributions provides $P(\theta | S=1)$ instead of $P(S=1 | \theta)$. The necessary conversion can be done applying Bayes Theorem:

$$
P(S=1 | \theta) = \frac{P(\theta | S=1)\cdot P(S=1)}{P(\theta | S=1) + P(\theta | S=0)}
$$

\noindent
where

$$
\frac{P(\theta | S=1)}{P(\theta | S=1) + P(\theta | S=0)} = \frac{P(\theta | S=1)}{P(\theta)}
$$

\noindent
can be interpreted as a factor of how much more likely it is to observe at least one seizure in that phase of the cycle, compared to the prior of observing at least one seizure in general. When this factor is larger than one, it means that we are observing a phase where it is more likely to have seizures, i.e., seizures tend to concentrate around that phase of the cycle.

To combine the information from different significant cycles, this interpretation is leveraged to pose it as the combination of such factors. In practice, the geometric mean of these factors was computed, that way combining the constructive and destructive contributions of each cycle's information. The computation used the following formula:

$$
factor = \left(\prod_{i=1}^{N_{cycles}} \frac{P(\theta_i | S=1)}{P(\theta_i | S=1) + P(\theta_i | S=0)}\right)^{\frac{1}{N_{cycles}}}
$$
$$
P(S=1 | \theta_1, ..., \theta_{N_{cycles}}) = factor \cdot P(S=1)
$$

\noindent
where $\theta_i, i={1, ..., N_{cycles}}$ is the corresponding phase for the $i$-th significant cycle.

\section{Derivation of Periodicity-Aware LR}
\label{app:LR}

The \gls{LR} model outputs the probability $P(S=1|X)$, where $X$ represents the feature vector. As demonstrated by Karoly et al. \cite{Karoly2017CircadianProfile}, the following relationship holds:

$$P(S=1|X) = \frac{1}{1+exp(-(w_0 + \sum_{i}^{N}w_i X_i))} \Rightarrow log(odds) = log\left(\frac{P(S=1|X)}{P(S=0|X)}\right) = w_0 + \sum_{i}^{N}w_i X_i$$

\noindent 
where $N$ is the number of features, $w_0$ denotes the bias term (or intercept), and $w_i$ represent the model weights associated with the corresponding features $X_i$.

As such, in the ensemble approach, this can be leveraged to update the priors ($P(S=1)$) in accordance to the periodicity-aware estimate of the von Mises ($P'(S=1)$) by directly updating the intercept: 

$$w'_0 = w_0 - log\left(\frac{P(S=1)}{1-P(S=1)}\frac{1-P'(S=1)}{P'(S=1)}\right)$$

\noindent
where the periodicity-aware estimate of the von Mises ($P'(S=1)$) is the result of the combination of all significant cycles and is obtained as described in Appendix \ref{app:vonMises}. 

Both models (standard and ensemble) were implemented using L2 regularization, optimized with the LBFGS (Limited-memory BFGS) solver, and a regularization strength of 1.0. No class balancing strategy was used. 

\end{document}